\documentclass[12pt]{article}
\usepackage{fullpage}
\usepackage{epsf}
\usepackage{epsfig}
\usepackage{amsthm}
\usepackage{amsfonts}
\usepackage{color}
\usepackage{amsmath}
\usepackage{natbib}

\begin{document}

\newcommand{\bm}[1]{\mbox{\boldmath $#1$}}
\newcommand{\mb}[1]{\mathbf{#1}}
\newcommand{\bE}[0]{\mathbb{E}}
\newcommand{\bP}[0]{\mathbb{P}}
\newcommand{\ve}[0]{\varepsilon}
\newcommand{\mN}[0]{\mathcal{N}}
\newcommand{\iidsim}[0]{\stackrel{\mathrm{iid}}{\sim}}
\newcommand{\NA}[0]{{\tt NA}}

\title{\vspace{-1.5cm} Gaussian process single-index models as emulators
for computer experiments}
\author{Robert B.~Gramacy\\
  Booth School of Business\\
  University of Chicago\\Chicago, USA\\
  {\tt rbgramacy@chicagobooth.edu} \and
  Heng Lian\\
  Division of Mathematical Sciences\\
  Nanyang Technological University\\Singapore\\
 {\tt henglian@ntu.edu.sg}}
\date{}

\maketitle

\begin{abstract}
  A single-index model (SIM) provides for parsimonious
  multi-dimensional nonlinear regression by combining parametric
  (linear) projection with univariate nonparametric (non-linear)
  regression models.  We show that a particular Gaussian process (GP)
  formulation is simple to work with and ideal as an emulator for some
  types of computer experiment as it can outperform the canonical
  separable GP regression model commonly used in this setting.  Our
  contribution focuses on drastically simplifying, re-interpreting,
  and then generalizing a recently proposed fully Bayesian GP-SIM
  combination.  Favorable performance is illustrated on synthetic data
  and a real-data computer experiment.  Two {\sf R} packages, both
  released on CRAN, have been augmented to facilitate inference under
  our proposed model(s).

  \bigskip
  \noindent {\bf Key words:} surrogate model, nonparametric
  regression, projection pursuit
\end{abstract}

\section{Introduction}

A single-index model (SIM) is a linear regression model with a
univariate nonparametric link function.  It provides a parsimonious
way to implement multivariate nonparametric regression.  Concretely,
the SIM is represented by $\bE\{Y_i|x_i\} = f(x_i^\top \beta)$, where
$\beta = (\beta_1, \dots, \beta_p)$ is called the {\em index vector}
for a $p$--dimensional predictor variable $x_i$, and $f$ is called the
{\em link}.  The product $x_i^\top \beta$ is the {\em index} of the
$i^{\mathrm{th}}$ response, $Y_i$.  The parameters $\beta$ and $f$
(which is usually infinite dimensional) are estimated jointly.  SIMs
like these, although with random Gaussian predictors, were first
formulated by \cite{brill:1977,brill:1982}. They have since been
applied widely in areas such as econometrics and psychometrics
\citep{ichimura:1993}.

SIMs are a special case of projection pursuit regression
\citep[PPR,][Section
11.2]{fried:stuetzle:1981,hastie:tibsh:fried:2001}, which uses $M$
projections and links (called {\em ridge functions}): $\bE\{Y_i|x_i\}
= \sum_{m=1}^M f_m(x_i^\top \beta_m)$. Although increasing $M$
provides greater flexibility, estimation of a nonparametric $f_m$ can
become difficult.  Furthermore, inference can be ad-hoc, involving
layers of greedy forward steps, backfitting and cross validation (CV).
Although the resulting fit may indeed yield high predictive power, it
is often uninterpretable.  As a result, many authors actually prefer
SIM models because the inference is simpler, the interpretation is
straightforward, and the properties of the estimators are well
understood.  

This paper promotes a particular SIM as an {\em emulator} for computer
experiments \citep{sant:will:notz:2003}.  An emulator is a
nonparametric nonlinear predictor for the output $Y$ of computer
simulations run at a design of input configurations $X$.  The typical
choice is a Gaussian process regression (GPR) model.  Emulator
parameters (e.g., those for a GPR) often have no
interpretation. Predictions are made by integrating over their
posterior distribution, so that decisions based on them incorporate
uncertainty in the model fit.  Examples include sequential design by
active learning \citep{gra:lee:2009}, calibration
\citep{higdon:etal:2004}, and optimization
\citep{gramacy:taddy:2010}.\footnote{These extend earlier versions by
  \cite{seo00}, \cite{kennedy:ohagan:2001} and
  \cite{jones:schonlau:welch:1998}, respectively, which used point
  estimates to make the fit.}  One reason for promoting SIMs as
emulators is that considerable insights may be gleaned by directly
studying the ``nuisance'' quantities, comprising of a projection and
indices.

Most of the literature on SIM inference is frequentist---see
\cite{anto:greg:mck:2004} for a nice review.  Until very recently, the
limited Bayesian work on SIMs focused on using splines for the link,
$f$ \citep{anto:greg:mck:2004,wang:2009}.  The impact of that work is
two-fold.  It represents a new direction in inference for SIM models
which, as illustrated empirically, can offer improvements over the
classical approach.  It also suggests a simple way that (Bayesian)
spline models, which are widely used for univariate nonparametric
regression, can be used in the multivariate setting where their
successes has been far more limited.

\cite{choi:shi:wang:2011} suggested that a GP prior be placed on $f$,
leading to a so-called GP-SIM.  The motivation and impact of such a
model is not immediately clear.  Unlike splines, GPRs already scale
naturally from one to arbitrary dimensions without projection.  So
this poses the question: what is gained from the GP-SIM approach
(either in the SIM context or as an emulator for computer experiments
where direct GPRs are commonplace)?  \cite{choi:shi:wang:2011} showed,
by simulation, that the GP-SIM reduce predictive error over the
canonical GPR (with an isotropic correlation function) in some cases,
e.g., on synthetic data generated within the SIM class, and on a
standard real-data benchmark.  However the MCMC algorithm is much more
complicated than SIM with splines or the canonical GPR.

Much of this extra effort is unnecessary.  There is a simpler
formulation of the GP-SIM which is comparable in computational
complexity to both SIM with splines and the canonical GPR, and works
just as well if not better.  Our primary contribution lies in
communicating this reformulation and its benefits, and promoting the
(new) GP-SIM as an emulator for computer experiments.  We contend that
the new formulation is easier to implement and portable to more exotic
modeling frameworks. 
In this way our work can be seen as a honing of the
\cite{choi:shi:wang:2011} approach, and as a subsequent generalization
and application to an important class of multivariate nonparametric
regression problems, namely computer experiments.  Finally, we provide
software as modifications to two existing {\sf R} \citep{R} packages
available on CRAN.  Together these support every feature discussed
herein, including all but one suggested extension.

The remainder of the paper is organized as follows.  In Section
\ref{sec:hier} we review the GP-SIM hierarchical model formulation of
\cite{choi:shi:wang:2011}.  We then introduce our simplifications, a
reformulated model, and a more efficient Monte Carlo inference method.
In Section \ref{sec:example} we provide some illustrative and
comparative examples on synthetic data, essentially extending the
\cite{choi:shi:wang:2011} results (with the new formulation) to
include a comparison to GPRs with a separable correlation---a stronger
straw man.  In Section \ref{sec:comp} we
provide a detailed example of a real-data computer experiment
involving computational fluid dynamics simulations of a rocket
booster. 
We conclude in Section \ref{sec:discuss} with a discussion of
extensions including a worked example of a treed version of the GP-SIM
on the rocket booster experiment, and applications to sequential
design, optimization, and classification.

\section{Hierarchical model and MCMC inference}
\label{sec:hier}

\subsection{Original formulation}

The Bayesian hierarchical formulation of the GP-SIM described by
\cite{choi:shi:wang:2011} is provided below. We have changed the
presentation/notation from its original version to make some
particular points more transparent and to ease the transition to our
new formulation.
\begin{align}
Y_i|x_i, \beta, f, \tau^2 & \iidsim \mathcal{N}(f(x_i^\top \beta), \tau^2), \; \mbox{ for }
\; i=1,\dots,n, \nonumber \\
\beta &\sim \Pi, \; \mbox{ subject to } \; ||\beta|| = \beta^\top \beta
=1, \;\;\; 
\tau^2 \sim \mathrm{IG}(a_\tau/2, b_\tau/2) \label{eq:hier} \\
f|\theta, \sigma^2 &\sim \mathrm{GP}(\theta, \sigma^2) \; \equiv\;
f(X\beta|\theta, \sigma^2)
\sim \mathcal{N}_n(0,
\sigma^2 K(X\beta;\theta)), \nonumber \\
\theta &\sim \mathrm{G}(a_\theta, b_\theta) \;\;\;\;\; \mbox{and} 
\;\;\;\;\; 
\log(\sigma) \sim \mathcal{N}(-1,1) \nonumber
\end{align}
G is the gamma distribution, IG the inverse-gamma distribution,
$\mathcal{N}$ the normal distribution, and $a_\tau$, $b_\tau$,
$a_\theta$, $b_\theta$ are known constants.  The rows of the design
matrix $X$ contain $x_1^\top, \dots x_n^\top$.  The prior
distributions for $f$ and $\beta$ require some explanation.

For $f$ notice the lack of explicit conditioning on $X\beta$ even
though its presence is helpful for thinking about the relevant finite
dimensional distributions.  The equivalence ($\equiv$) illustrates
what the GP prior implies when $X\beta$ is supplied as an argument.
In shorthand we have $f_n \equiv [f(x_1^\top \beta), \dots,
f(x_n^\top\beta)]^\top \sim \mathcal{N}_n(0, \sigma^2K_n)$, i.e., a
$n$-variate normal prior distribution with zero mean and $n \times n$
correlation matrix $K_n \equiv K(X\beta ; \theta)$ which has
$(i,j)^{\mathrm{th}}$ entry
\begin{equation}
K(x_i^\top \beta, x_j^\top \beta; \theta) = \exp \left\{
-\frac{(x_i^\top \beta - x_j^\top \beta)^2}{\theta}  \right\}.
\label{eq:K}
\end{equation}
In other words, the GP prior uses a Gaussian correlation function,
with {\em length-scale} (or {\em range}) parameter $\theta$, applied
to the projected indices $X\beta$.

For $\beta$ the important part isn't the choice of prior distribution
$\Pi$, but rather the constraint that it lies on the unit $p$-sphere.
The explanation is that we are jointly modeling $f$ and $\beta$ which
interact as $f(X\beta)$, so $\beta$ is only identifiable up to a
constant of proportionality.  
In practice $\Pi$ is chosen to be uniform, but a more flexible
Fisher--von Mises prior distribution (which has a built in unit-sphere
constraint) can be used if prior information is available.  Observe
that the model leaves the sign of $\beta$ unidentified since $-\beta$
leads to the same likelihood under either specification---an issue
that is glossed over by previous Bayesian treatments of SIMs.
Importantly, there is posterior consistency of $\beta$ under this
setup \citep{choi:shi:wang:2011}.

\subsubsection*{Inference by Monte Carlo}

Sampling from the posterior distribution of the parameters proceeds by
Markov chain Monte Carlo (MCMC), specifically by
Metropolis--within--Gibbs by iterating through full conditionals for
$f_n$, $\sigma^2$, $\beta$, $\theta$, then $\tau^2$.  The actual
expressions for the conditionals are not reproduced here.  The first
two are standard distributions ($\mathcal{N}_n$ and IG, respectively)
yielding Gibbs updates, whereas the latter three require
Metropolis--Hastings (MH).  

The $\beta$ parameter is treated as a single block, and the proposals
come in a random--walk (RW) fashion using a Fisher--von Mises
distribution whose modal parameter is set to the previous value
\citep{anto:greg:mck:2004}.  We add that, in this context, the lack of
identifiability of the sign of $\beta$ is akin to the label-switching
problem for MCMC inference of (Bayesian) mixture models.  In Appendix
\ref{a:sign} we offer some post-processing suggestions for resolving
the ``labels'' (signs) of the two ``clusters'' (modes of the marginal
posterior posterior for $\beta$).  We presume that $\theta$ and
$\sigma^2$ use RW-MH, but \cite{choi:shi:wang:2011} do not provide
these details.  
Sampling $f_n|\theta, \sigma^2$, a vector of $n$ latent variables,
proceeds via the well-known {\em kriging} equations discussed in more
detail for the reformulated model to follow.

\subsection{Reformulation}

Our first important observation has to do with the nature of (the lack
of) identifiability of $\beta$ without unit ball restriction.  In the
particular case of a GP prior for $f$, through $K_n$ in
Eq.~(\ref{eq:K}), it is easy to see why it can only be identified up
to a constant.  The model is over parameterized.  The quantity
$\beta/\sqrt{\theta}$ is identifiable (up to a sign).  Or, in other
words, if we remove $\theta$ from the model (or fix it to one) and
free $\beta$ from its unit-sphere restriction, then $\beta$ is
identifiable up to a single sign, $-\beta$ or $\beta$.  In many cases,
like in computer experiments (and mixture models), it does not matter
at all (as long as the MCMC explores all possibilities) since $\beta$
is not of direct interest.  In others, e.g., where variable selection
is important \citep{wang:2009}, inability to identify signs poses no
real issue.  When inference for $\beta$ {\em is} a primary goal, then
identifiability is, of course, important.  However, we show how some
simple heuristics for reconciling the signs [Appendix \ref{a:sign}]
does allow extra explanatory power that is uncanny in this context,
through the indices $X\beta$.

Supposing we effectively remove $\theta$ from the model, our second
important observation is that $\beta$ ought to be treated as a
parameter to the correlation function (\ref{eq:K}), which otherwise
would have none left. Our proposal is to re-interpret the correlation
function as
\begin{equation}
K^*(x_i,
x_j;\beta) = \exp\left\{-(x_i^\top \beta - x_j\beta)^2 \right\} =
\exp\{- (x_i - x_j)^\top \beta \beta^\top (x_i - x_j) \}. \label{eq:Kstar}
\end{equation}
This is a special rank-1 case of a Gaussian correlation structure with
an inverse length-scale {\em matrix} $\Sigma = \beta \beta^\top$.  So
we have a convenient re-interpretation of the GP-SIM.  It is just a
canonical GPR model with an odd correlation function.  Therefore, an
equivalent hierarchical model to (\ref{eq:hier}), combining the GP
prior and the IID normal likelihood are combined into a single
expression for $Y = (Y_1, \dots, Y_n)$, is
\begin{align}
  Y|X, \beta, \sigma^2, \eta &\sim \mathcal{N}_n(0, \sigma^2 K_n), 
  \;\; \mbox{ with } 
   K_n \equiv K(X;\beta, \eta) \label{eq:hier2}  \\
  \beta &\sim \Pi, \;\;\;  
 \sigma^2 \sim \mathrm{IG}(a_\sigma/2, b_\sigma/2), \;\;\; 
  \eta \sim \mathrm{G}(a_\eta, b_\eta). \nonumber
\end{align}
$K(X;\beta, \eta)$ is the nugget-augmented correlation function
$K(x_i, x_j ; \beta, \eta) = K^*(x_i, x_j | \beta) +
\eta_{\delta_{i,j}}$, where $K^*$ is given in Eq.~(\ref{eq:Kstar}).
The parameter $\eta$ is called the nugget parameter.  This
nugget-augmented correlation is known \citep[e.g.,][Appendix
B]{gramacy:2005} to be equivalent to one having two separate variance
terms ($\tau^2, \sigma^2$) since $\eta \equiv \tau^2/\sigma^2$ and is
preferable when using MCMC, as described below.

Now the first line in Eq.~(\ref{eq:hier2}) is equivalent to the $Y_i$
and $f$ lines in Eq.~(\ref{eq:hier}).  The rest, i.e., the prior
distribution s, are slightly different.  We are free to choose
whatever prior distribution, $\Pi$, we wish for $\beta$, but note that
a Fisher--von Mises prior distribution would not be recommended
because this would severely constrain the new model.  If prior
information is available, then any sensible way of encoding it
suffices.  Our default is $|\beta_j| \iidsim \mathrm{G}(a_\beta,
b_\beta)$ for particular choices of $a_\beta$ and $b_\beta$, however
multivariate normal (MVN) prior distributions on $\beta$ may be
sensible when prior information is available.  In either case, we
assume that the design matrix $X$ has been pre-scaled to lie in the
unit $p$-cube.  This makes choosing sensible defaults for $a_\beta$,
$b_\beta$, $a_\eta$, and $b_\eta$ much easier [see Section
\ref{sec:example}].

The benefits of this new formulation may not, yet, be readily
apparent.  There is only one fewer parameter.  But we can obtain more
efficient inference by MCMC since we have eliminated $O(n)$ latent
variables, $f_n$.  Our setup suggests implementing the GP-SIM as a GPR
with a new correlation function, so its implementation (given existing
GPR code) is trivial, and thus is ripe for extension [see Section
\ref{sec:discuss}].  Before turning to details of inference and
implementation we remark that the posterior consistency result
provided by \cite{choi:shi:wang:2011} applies in our reformulated
version.  To see why, observe that any continuous distribution on
$\beta$ can be decomposed into a distribution on $||\beta||$ and
another on $\beta/||\beta||$.  For example, $\beta \sim N(0, I)$ is
the same as $||\beta||^2 \sim \chi_p^2$ and $\beta / ||\beta||$
uniform on the sphere.  In short, the models are essentially
identical, and so are their properties.  We prefer slightly different
prior distributions (mainly because of defaults in existing GP
software), and these do not materially change the nature of the
posterior distributions.  The key is that our re-interpretation allows
for a simpler inferential approach.

\subsubsection*{Inference by Monte Carlo}

An advantage of the nugget-augmented correlation function is that the
$\sigma^2$ parameter can be integrated out analytically in the
posterior distribution.  This means it is never needed directly, even
in the predictive equations to follow.  We obtain the following
marginalized conditional posterior distribution for any $K$, which in
our particular GP-SIM case is $K(X; \beta, \eta)$:
\begin{equation}
  p(K|X,Y) = p(K) \times \frac{(b_\sigma/2)^{\frac{a_\sigma}{2}}
    \Gamma[(a_\sigma+n)/2]}{|K_n|^{\frac{1}{2}}(2\pi)^{\frac{n}{2}}\Gamma[a_\sigma/2]} \times
  \left(\frac{b_\sigma +
      Y K_n^{-1} Y}{2}\right)^{-\frac{a_\sigma+n}{2}}.  \label{eq:gpk}
\end{equation}
See \citet[][Appendix A.2]{gramacy:2005} for a full derivation in a
slightly more general setup.  The quantity $p(K)$ is a stand-in for
$\Pi(\beta) \times \mathrm{G}(\eta;a_\eta, b_\eta)$ and $K_n$ is the
$n \times n$ matrix implied by $K(X; \beta, \eta)$.  The Jeffreys'
prior $p(\sigma^2) \propto 1/\sigma^2$, i.e., choosing $(a_\sigma,
b_\sigma) = (0,0)$, is preferred when there is no prior information
about the scale of covariance.  It may be used as long as $n > 1$ and
leads to a simplified expression for $p(K|X,Y)$ upon taking $\Gamma[0]
\equiv 1$ and $0/0\equiv 1$.

The significance of this result is that MCMC need only be performed
for $K$ via $\beta$ and $\eta$.  So we only need establish a chain for
$p+1$ parameters compared to $n + p + 3$ parameters in the original
formulation.  The time required for each MH round is unchanged
relative to the original formulation at $O(n^3)$ to decompose $K_n$
for each newly proposed $\beta$.  However, by avoiding the unnecessary
sampling of $n$ latents in each round, which is $O(n^2)$ [see
Eqs.~(\ref{eq:predgp}--\ref{eq:preds2}) below], we not only save
(slightly) on computational cost, but also (significantly) reduce the
Monte Carlo error of the MCMC by having a lower dimensional chain [see
Section \ref{sec:synth}].

In the remainder of this section we outline how our MCMC scheme
further deviates from \citet{choi:shi:wang:2011}, and comment on some
computational advantages that are available in our setup.  We start
with the nugget $\eta$, which requires MH.  A good RW proposal is the
positive uniform sliding window $\eta' \sim \mathrm{Unif}[3\eta/4,
4\eta/3]$.  The proposed $\eta'$ may be accepted or rejected according
to a ratio involving the proposal probabilities and
$p(K'|X,Y)/P(K|X,Y)$ with $\beta$ fixed, i.e., implementing
Metropolis-within-Gibbs sampling.

Drawing $\beta$ for fixed $\eta$ can proceed similarly given a
suitable proposal for $\beta$.  As components of $\beta$ can be highly
correlated [see Section \ref{sec:synth}], we prefer to update $\beta$
in a single block.  Assuming the support of the prior distribution
$\Pi$ for $\beta$ is $\mathbb{R}^p$, a RW-MVN proposal centered at
$\beta$ is a reasonable choice, i.e., $\beta' \sim N_p(\beta,
\Sigma_\beta)$.  If posterior correlation information about $\beta$ is
known, e.g., from a pilot MCMC run, this can be used to inform a good
choice of $\Sigma_\beta$ which can be crucial for obtaining good
mixing in the Markov chain.  Such tuning would be much harder with
Fisher--von-Mises distributions in the setup of
\cite{choi:shi:wang:2011}.  Note that it helps to reconcile the signs,
or ``labels'', of $\beta$s sampled from the posterior distribution
[Appendix \ref{a:sign}] before using them to estimate $\Sigma_\beta$.
For the pilot run, or otherwise, we find that $\Sigma_p =
\mathrm{diag}_p(0.2)$ works well when $X$ is pre-scaled to lie in
$[0,1]^p$.

If an application demands that all/both ``labels'' of $\beta$ be
explored {\em a posteriori}, then we suggest using the signs of
$\mathcal{N}_p(0, \Sigma_\beta)$ to create a compound proposal: take
$\beta' = s * b$, a component-wise product where $s \sim
\mathrm{sign}[\mathcal{N}_p(0, \Sigma_\beta)]$ and $b \sim
\mathcal{N}_p(\beta, \Sigma_\beta)$. Our experience is that such
random sign changes only slightly alter the MH acceptance ratio when
$\Sigma_\beta$ is tuned from a pilot run.  The reason is that ours is
a variation on a scheme that periodically tries $\beta' = -\beta$,
which always accepts. Observe that proposing $s \sim
\mathrm{sign}[\mathcal{N}_p(0, \Sigma_\beta)]$ requires calculating
MVN orthant probabilities for the MH ratio.  For this we recommend the
method of \citet{miwa:hayter:kuriki:2003} as implemented by
\citet{craig:2008}.

Sampling from the posterior predictive distribution is easy given a
collection of $\beta$ and $\eta$ values.  The distribution of $Y(x^*)$
given $X$, $Y$, and $K$ is Student-$t$ with
\begin{align}
  \mbox{mean} && \hat{y}(x|X,Y,K) &= k_n^\top(x)
  K_n^{-1}Y,\label{eq:predgp} \\
\mbox{scale} &&
 \hat{\sigma}^2(x|X,Y,K) &=
\frac{(b_\sigma + Y K_n^{-1} Y) [K(x, x) - k_n^\top(x)K_n^{-1} k_n(x)]}{a_\sigma + \hat{\nu}},
\label{eq:preds2}
\end{align}
and degrees of freedom $\hat{\nu} = n-1$,
where $k_n^\top(x)$ is the $n$-vector whose $i^{\mbox{\tiny th}}$
component is $K(x,x_i; \beta, \eta)$.  These equations, which are a
minor extension of the classic kriging equations, are extremely
versatile.  They can be used to obtain samples from the posterior
predictive distribution; to obtain average mean-and-quantile
posterior summaries; or as a basis for sequential design [see Section
\ref{sec:doe}].  We can even sample {\em jointly} from the predictive
at a design of multiple new locations $X^*$ to obtain predictive {\em
  sample paths} via a multivariate Student-$t$ distribution derived by
simple matrix extensions of Eq.~(\ref{eq:predgp}--\ref{eq:preds2}).
All of these would be extremely difficult under the original
formulation.

We have so far been leveraging the GPR-only formulation of the GP-SIM
model, trying to forget its roots as an index model.  However, we can
still obtain a sample of the indices by simply collecting samples of
$x^{*\top} \beta$ at any location(s) $x^*$.  As illustrated in
Sections \ref{sec:synth} and \ref{sec:comp}, this can be useful for
assessing goodness of fit and add explanatory/interpretive power even
when aspects of these quantities are technically not identifiable.

\section{Implementation and illustration}
\label{sec:example}

Here we illustrate our implementation(s) of the re-formulated GP-SIM
and on synthetic data, turn to real data from a computer experiment in
Section \ref{sec:comp}.  We primarily follow \cite{choi:shi:wang:2011}
and compare GP-SIM to canonical GPRs, using isotropic and separable
Gaussian family correlation functions.  Specifically,
\begin{equation}
K(x_i, x_j|\theta, \eta) = \label{eq:K2}
 \exp\left\{ - \sum_{k=1}^{p}
  \frac{|x_{ik} - x_{jk}|^{2}}{\theta_{k}}\right\} + \eta_{\delta_{i,j}}.
\end{equation}
This is the separable case, where $\theta = (\theta_1, \dots,
\theta_p)$ allows different length-scale parameters in each dimension.
The isotropic case fixes $\theta_1 = \cdots = \theta_p$.
\cite{choi:shi:wang:2011} did not include a separable comparator,
which (as we shall see) is superior for multivariate regression.

All experiments used the {\tt tgp} package defaults for prior and
proposal distributions unless explicitly stated otherwise.  [Section
\ref{sec:discuss} contains the software details.] We did not propose
random sign changes for $\beta$.  Rather, we initialized the MCMC with
$\beta_j = 1/2$ and allowed the chain to explore the mode around one
of the two labels.  The results are very similar with a compound
proposal mechanism involving sign changes but requires a more careful
application of the methods described in the Appendix to reconcile the
signs and make intelligible diagnostic and descriptive plots like the
ones shown below.

Our quantitative metric of comparison between predictors is
Mahalanobis distance (Mah), as proposed by \citet{bastos:ohagan:2009}.
It is similar to RMSE but allows for covariance in the predicted
outputs to be taken into account.  For a vector of responses $y =
(y(x_1), \dots y(x_N))^\top$ at $N$ hold-out predictive locations, the
distance is given by $\mathrm{Mah(y; \mu, \Sigma)} = (y - \mu)^\top
\Sigma^{-1} (y - \mu)$, where $\mu$ and $\Sigma$ are estimates of the
predictive means and covariances for the $N$ locations $y$.  The
distance can be interpreted as an approximation to the (log) posterior
predictive probability of $y$.

\subsection{Synthetic data in the SIM class}
\label{sec:synth}

Consider data generated within the SIM class.  The function of the
index, $t$, is periodic:
\[
f(t) =  \sin\left(\frac{\pi t}{5}\right)
        + \frac{1}{5}\cos\left(\frac{4\pi t}{5}\right).
\]
The data are observed as $Y_i \sim \mathcal{N}(f(x_i^\top \beta),
0.1^2)$ with 4-dimensional predictors $x_i$ and $\beta = (2.85, 0.70,
0.99,-0.78)$.  In a Monte Carlo experiment we simulated design
matrices with $n=45$ rows uniformly in $[0,1]^4$, and then
conditionally sampled 45 responses thus forming our training set.  We
similarly simulated a testing design matrix of 200 rows, recording the
corresponding true (no-noise) response for making predictive
comparisons via Mahalanobis distance.
\begin{figure}[ht!]
\centering
\begin{minipage}{8cm}
\vspace{0.2cm}
\hspace{3.75cm} (a)\\
\includegraphics[trim=0 20 0 40,scale=0.55]{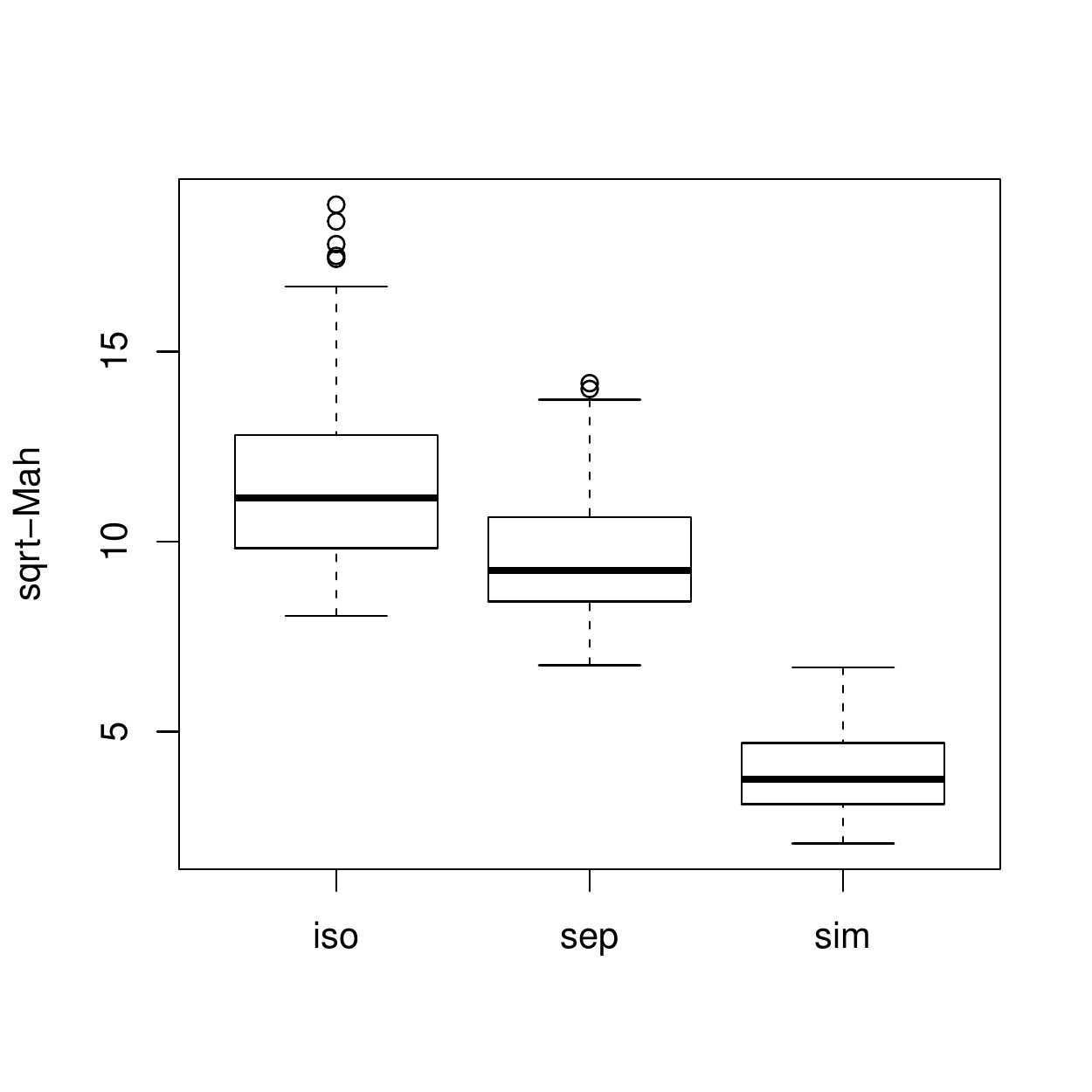}
\end{minipage}
\hspace{0.2cm}
\begin{tabular}{l|rrr}
(b) & \multicolumn{3}{c}{$\sqrt{\mathrm{Mah}}$} \\
&      iso       &       sep     &         sim        \\ 
\hline
 Min.    & 8.049   & 6.754   & 2.052   \\
 1st Qu. & 9.857   & 8.437   & 3.091  \\
 Median  & 11.149   & 9.251   & 3.746  \\
 Mean    & 11.569  & 9.611   & 3.899  \\
 3rd Qu. & 12.795  & 10.621   & 3.900  \\
 Max.    & 18.868   & 14.171   & 6.688  
\end{tabular}
\caption{Square-root Mahalanobis distance results for data generated
  in the SIM class in terms of boxplots (a) and a numerical
  summary (b).}
\label{f:sinrmse}
\end{figure}
This was repeated 100 times, each time fitting the three models in
question, sampling from their respective predictive distributions, and
calculating Mahalanobis distances.  The results are summarized in
Figure \ref{f:sinrmse}, and the ranking they imply is no surprise.
What is particularly noteworthy is the rarity of deviation from this
ordering in all 100 repetitions.  The GP-SIM {\em always} had a lower
Mahalanobis distance than the separable GPR which was itself better
than the isotropic GPR 83\% of the time.  In fact, observe that the
worst GP-SIM distance is better than the best of the other two.

\begin{figure}[ht!]
\centering
\vspace{0.5cm}
\hspace{-0.5cm}
\begin{minipage}{11.5cm}
\hspace{2.75cm} (a) \hspace{5cm} (b) \\
\includegraphics[scale=0.55,trim=2 68 20 45,clip=TRUE]{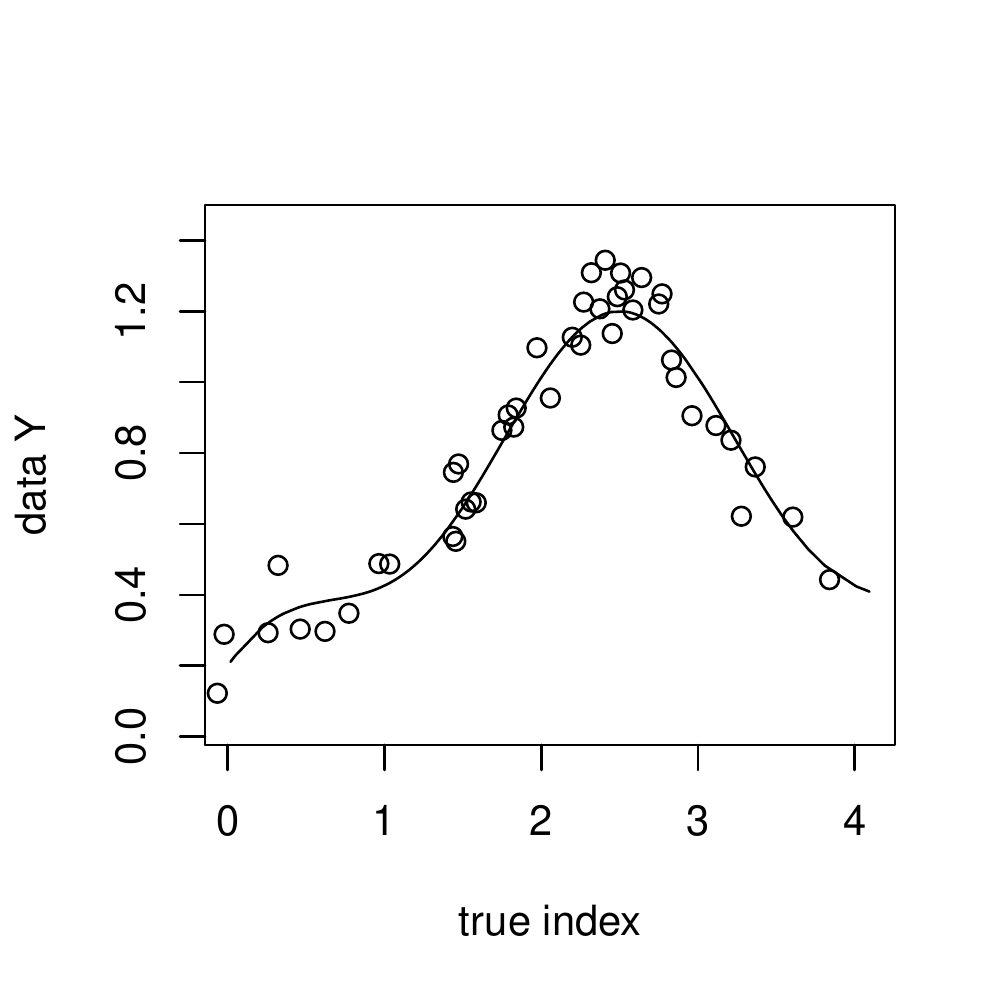}
\includegraphics[scale=0.55,trim=0 68 20 45,clip=TRUE]{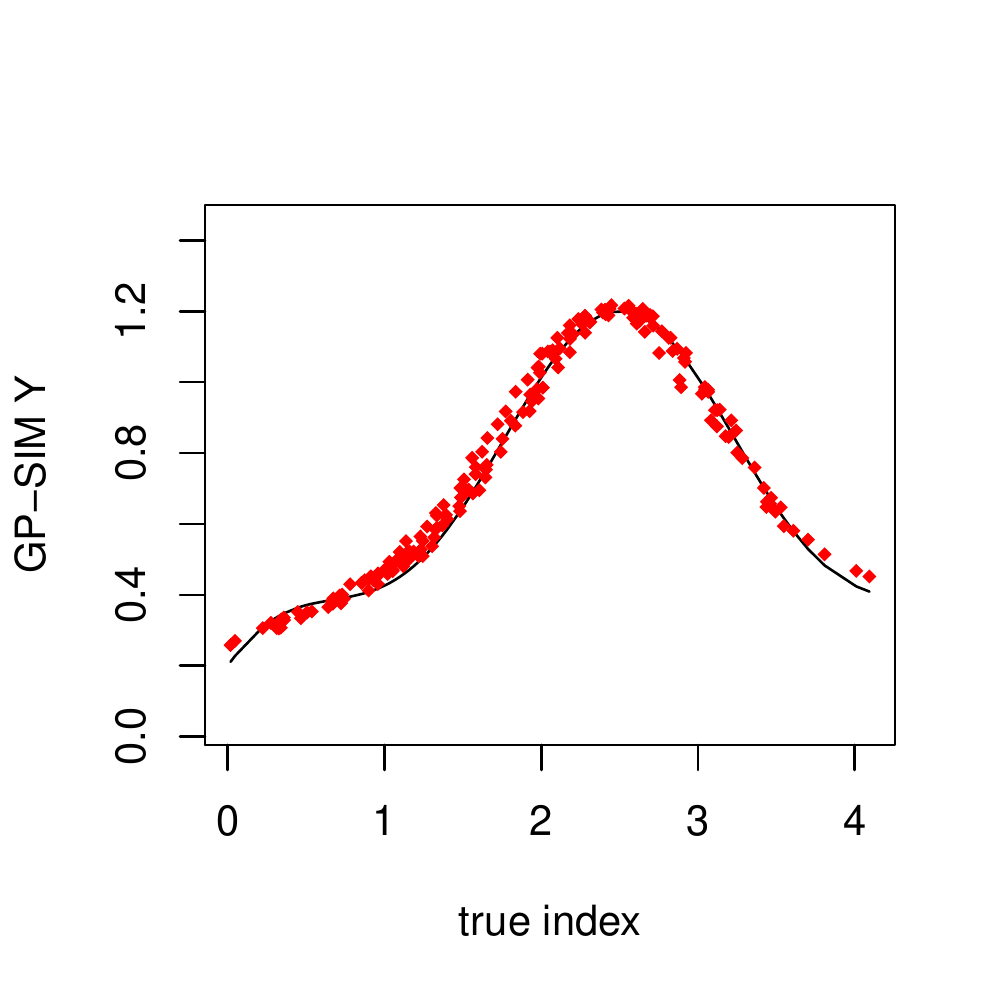}\\

\hspace{2.75cm} (c) \hspace{5cm} (d) \\
\includegraphics[scale=0.55,trim=2 0 20 45,clip=TRUE]{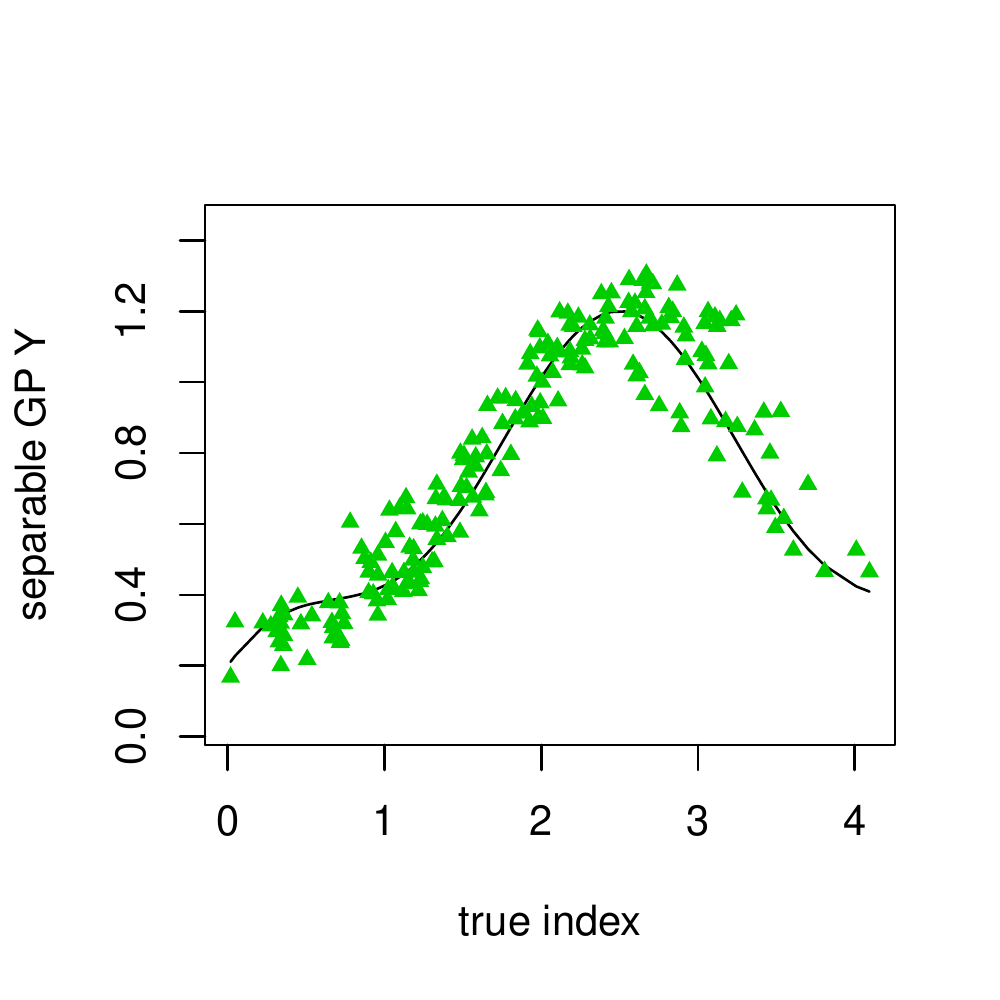}
\includegraphics[scale=0.55,trim=0 0 20 45,clip=TRUE]{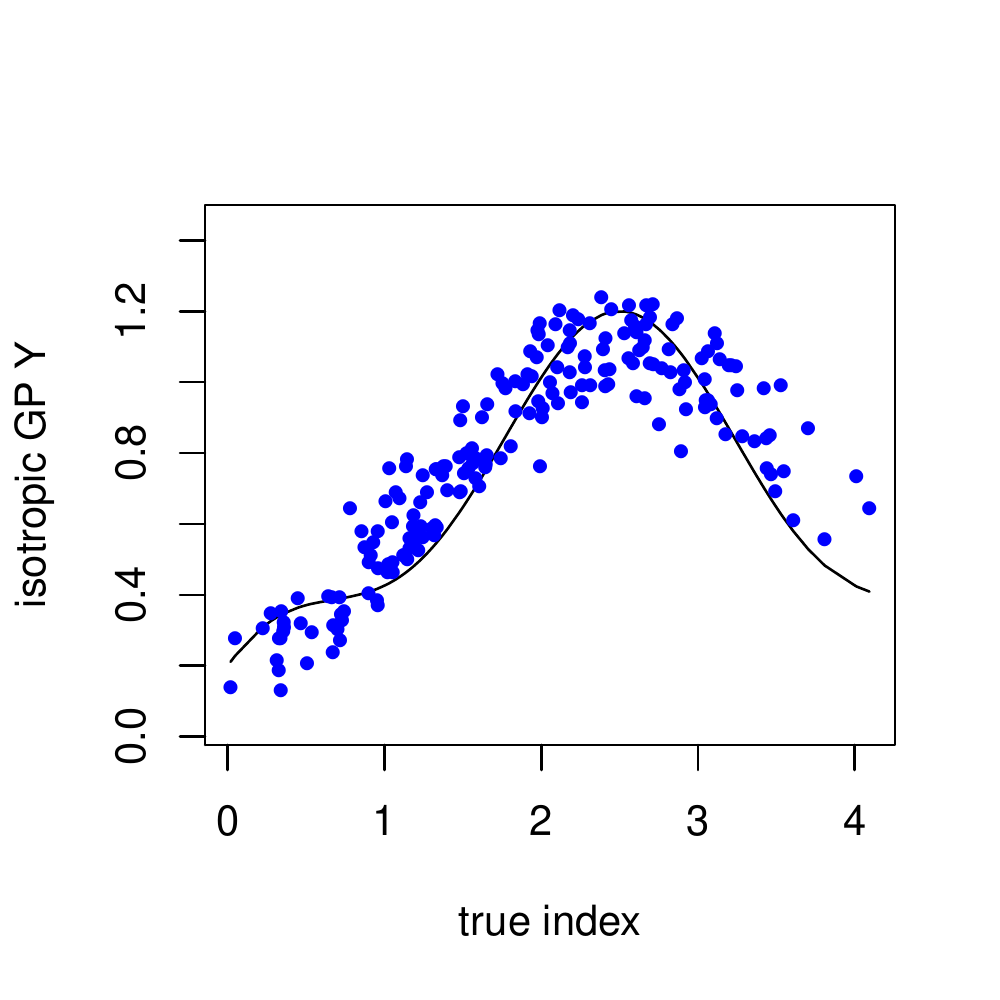}
\end{minipage}
\vline
\hspace{0.15cm}
\begin{minipage}{4cm}
\hspace{2cm} (e) \\
\includegraphics[scale=0.6,trim=55 0 20 45,clip=TRUE]{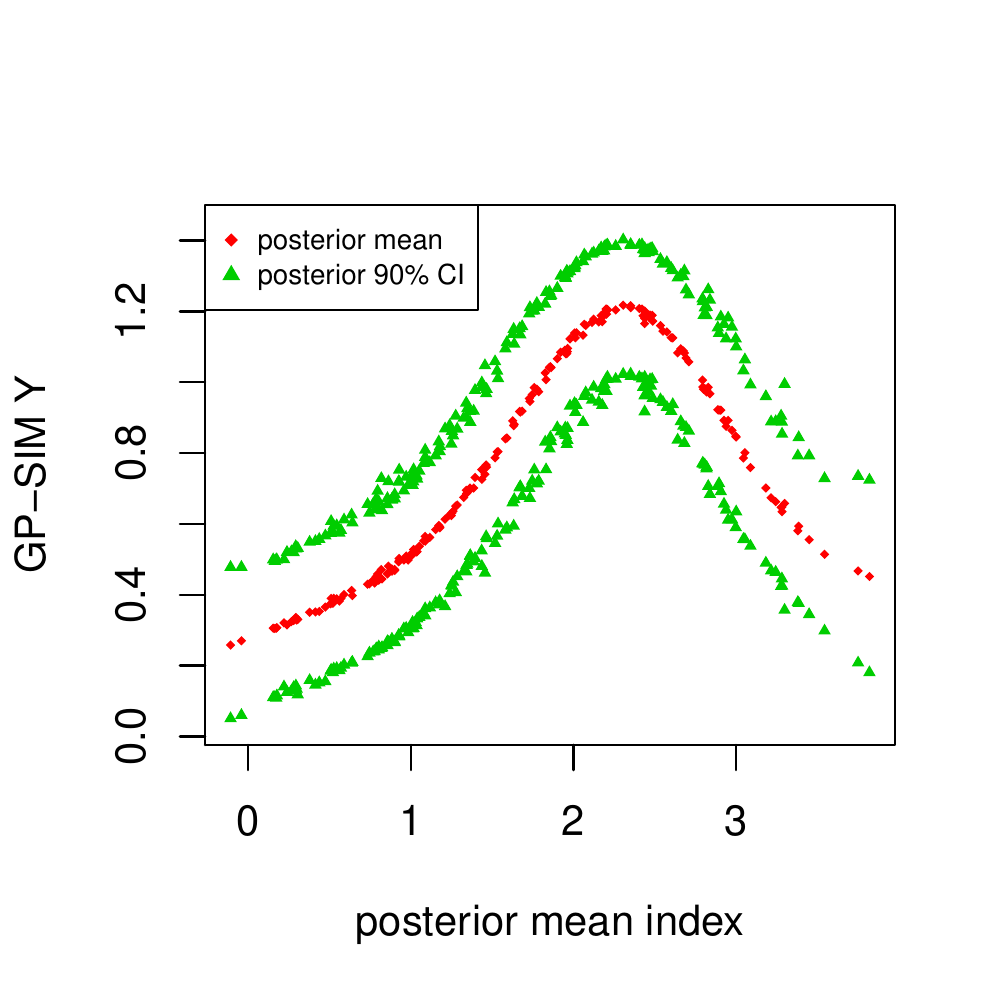}\\
\vfill
\end{minipage}
\caption{Training data (a) and posterior predictive mean values under
  the GP-SIM (b), separable GP (c), and isotropic GP (d) plotted
  versus the true index.  The solid line in each plot indicates the
  true index-response relationship.  Panel (e) plots the GP-SIM with
  estimated indices.}
\label{f:sinindex}
\end{figure}

Figure \ref{f:sinindex} (a)--(d) displays, for one realization, the
data, the true function, and predicted means for the 200 test points
as functions of the true index.  The increasing amount of scatter in
(b) to (d) clearly explains the performance ordering seen in Figure
\ref{f:sinrmse}.  We can also plot the predicted mean versus the
fitted (posterior mean) indices.  See Figure \ref{f:sinindex} (e).
The reduction in scatter between (b) and (e) is due to the change in
the horizontal axis.  By plotting against the posterior mean index in
(e), variation in predictions due to uncertainty in estimation of the
link is masked.  The posterior 95\% predictive credible interval (CI)
of the response, which is also plotted as a function of the estimated
mean indices, provides a look at the advantage of the fully Bayesian
approach.  Ideally, we would like to see the posterior uncertainty in
the indices in the plot as well, but variability on multiple axes is
hard to visualize (although some of the uncertainty in the posterior
mean indices can be seen from the horizontal jitter of the dots).
Instead, we prefer to show the variability of the indices implicitly
though posterior uncertainty in $\beta$.  But before doing so, some
comments on how we obtain the fitted indices are in order.

First, the unidentifiable sign of $\beta$ can cause lack of
identifiability in the sign of the indices.  This was easy to correct
since all of the signs of the components of $\beta$ that were sampled
by MCMC were the same---we were only exploring one mode---so none of
the heuristics from Appendix \ref{a:sign} were needed.  Therefore no
adjustments to the indices were needed either.  Second, since the
length-scale $\theta$ is built-in to $\beta$ the range of the indices
will have a different scale than the true indices.  This is harder to
fix, but it isn't really necessary---you can get a nice plot of the
index-versus-response relationship without adjusting the scale [see
Figure \ref{f:sinindex}(e)].  Incidentally, if the data-generating
(true) $\beta$ is not on the unit-sphere (and it is not in this
example) then the same scaling problem arises under the original model
formulation.

The posterior correlation matrix we obtained for $\beta$ is shown in
Table \ref{t:bcor}.
\begin{table}[ht!]
\centering
\begin{tabular}{r|rrrr}
 &          $\beta_1$     &    $\beta_2$  &  $\beta_3$   & $\beta_4$ \\
\hline
$\beta_1$ & $1 (0.461)$ & $0.397$ & $0.565$ & $-0.482$ \\
$\beta_2$ & $0.397$ & $1 (0.166)$ & $0.100$ & $-0.461$ \\
$\beta_3$ & $0.565$ & $0.100$ & $1 (0.135)$ & $-0.206$ \\
$\beta_4$ &$-0.482$ & $-0.461$ & $-0.206$ & $1 (0.109)$
\end{tabular}
\caption{Posterior correlation matrix for $\beta$, with variances in
parentheses.}
\label{t:bcor}
\end{table}
Non-negligible correlation between the components of $\beta$ means we
can expect to obtain lower MC error by using the above estimated
values as $\Sigma_\beta$ for future runs.  In a second run of the MCMC
(of equal length) using this new proposal covariance we obtained an
effective sample size \citep{kass:1998} of more than seven times that
of the original chain.  This improvement would not have been possible
with a Fisher--von Mises proposal.

\begin{figure}[ht!]
\centering
\vspace{0.2cm}
(a) \hspace{6cm} (b) \\
\includegraphics[trim=0 10 0 35,scale=0.45]{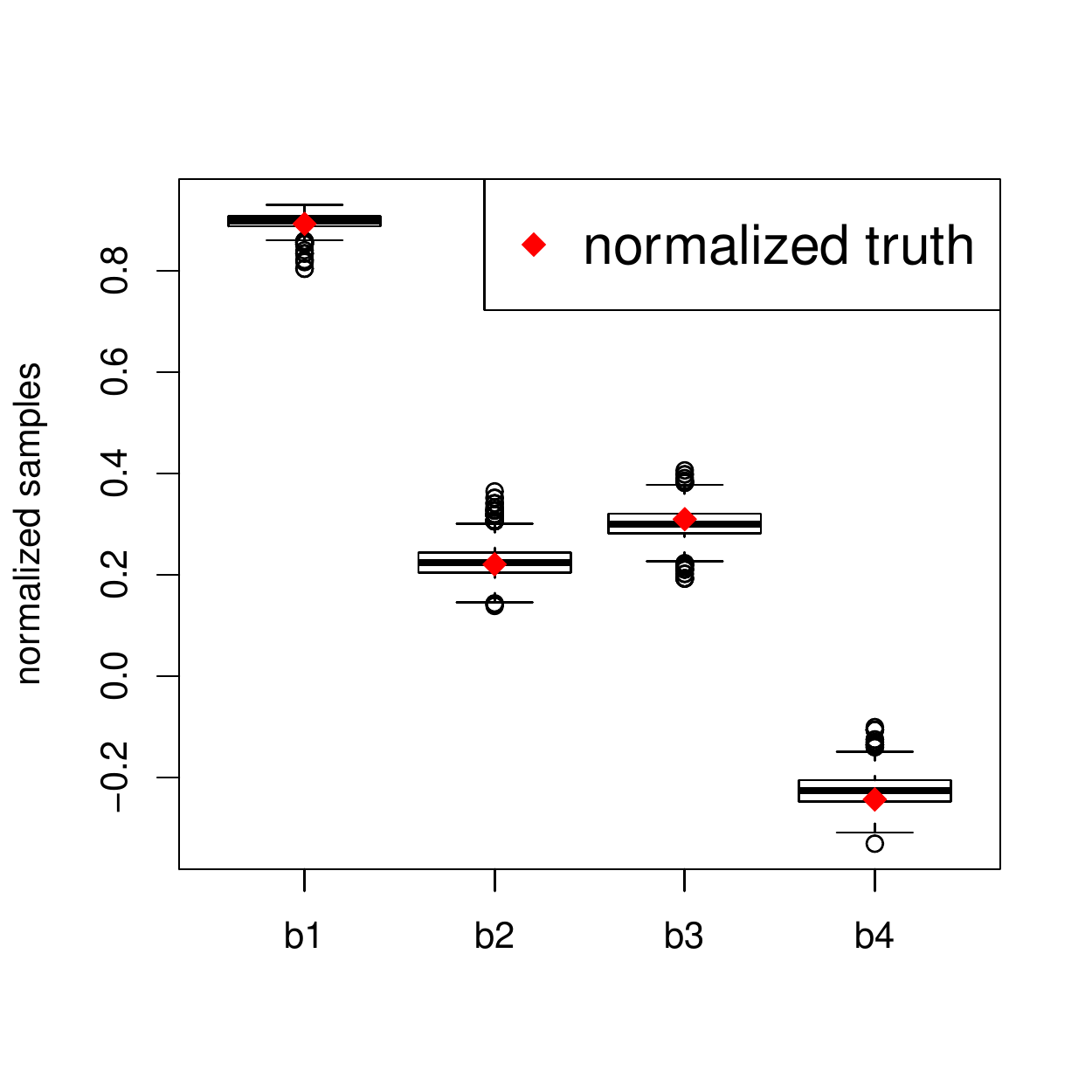}
\hspace{1cm}
\includegraphics[trim=0 10 0 35,scale=0.45]{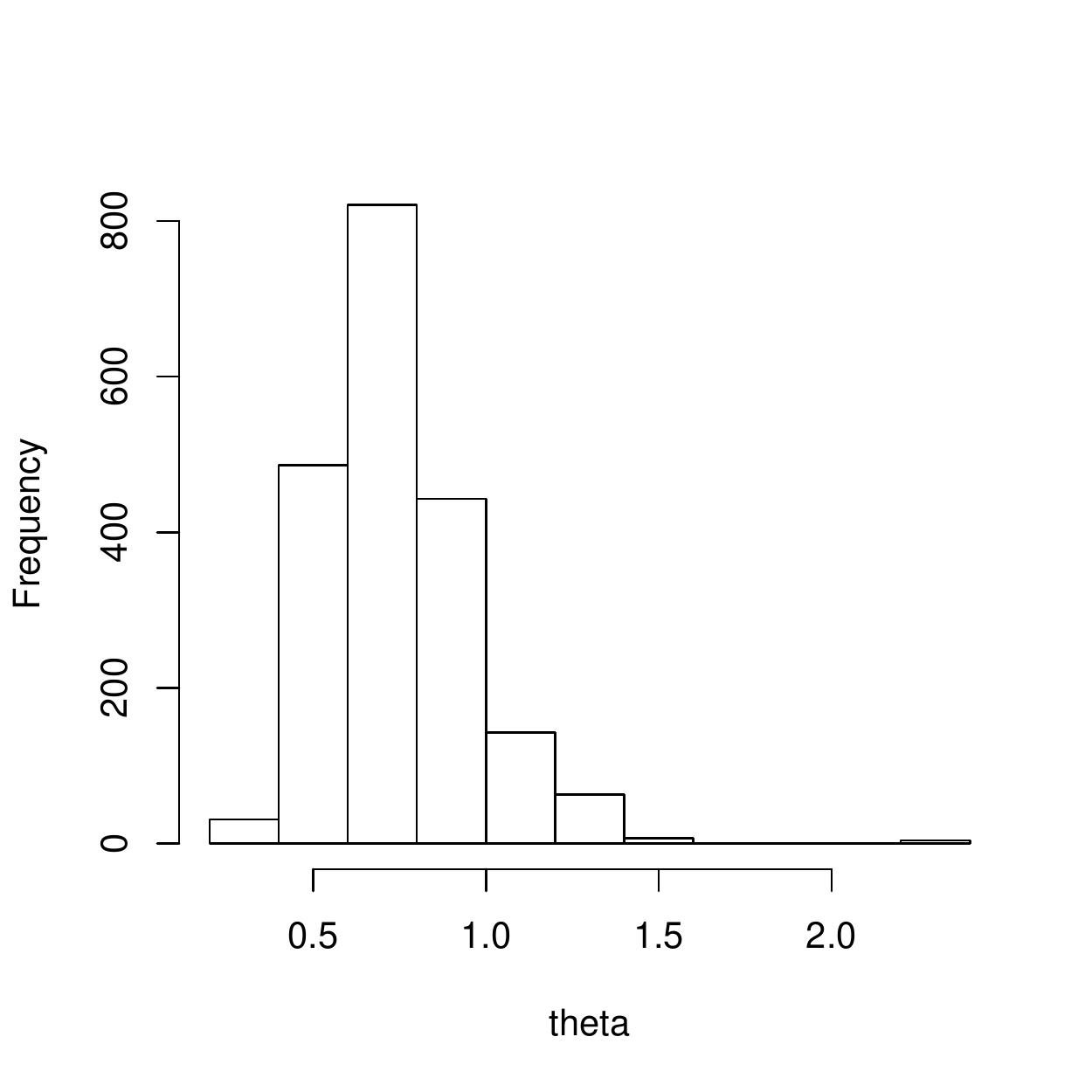}
\caption{Boxplots (a) collecting normalized samples from the posterior
  distribution for $\beta$, and a histogram (b) showing the implied
  values of $\theta$ based on the normalization.}
\label{f:betasamps}
\end{figure}
Figure \ref{f:betasamps}(a) shows the resulting samples, normalized
to lie on the unit sphere.  Observe that the posterior density is very
tight around the true (normalized) index vector.  The normalizing
constant of each $\beta$ sample implies a sample of the phantom
$\theta$ length-scale parameter via a reciprocal and square root.  A
histogram of these values is shown in panel (b).

\subsection{The borehole data: a deterministic function}
\label{sec:borehole}

An example that contrasts with the previous one is the borehole
function \citep{worley:1987}, as studied by many authors
\citep[e.g.][]{morris:mitchell:ylvisaker:1993}.  The response $y$ is
given by
\begin{equation}
y = \frac{2\pi T_u [H_u - H_l]}{\log\left(\frac{r}{r_w}\right) 
\left[1 + \frac{2 L T_u}{\log (r/r_w) r_w^2 K_w} + \frac{T_u}{T_l}
\right]}\,,
\label{eq:borehole}
\end{equation}
The eight inputs are constrained to lie in a rectangular domain:
\begin{align*}
r_w &\in [0.05, 0.15] & r &\in [100,5000] & T_u &\in [63070, 115600] &
T_l &\in [63.1, 116] \\
H_u &\in [990, 1110] & H_l &\in [700, 820] & L &\in [1120, 1680] & 
K_w &\in [9855, 12045].
\end{align*}
We offer some experimental results on data obtained from the borehole
function (without noise) in order to highlight how the GP-SIM compares
to other GPR models when the data-generating mechanism is far outside
the SIM class.  Approximating Eq.~(\ref{eq:borehole}) with a
transformation of a linear combination of the eight predictors will be
crude in comparison to many alternatives.  Note that we still fit the
GPRs/GP-SIM with a nugget even though the function is deterministic.
For an explanation, see \cite{gra:lee:2011}.

We generated a size 250 Latin hypercube design \citep[LHD,][Section
5.2.2]{sant:will:notz:2003} constrained to the above rectangle, and
obtained 250 responses, $y$.  A hold-out testing set of size 1000 is
similarly obtained, and after fitting the GP-SIM and canonical GPRs as
in Section \ref{sec:synth}, it is used to calculate Mahalanobis
distances to measure predictive accuracy.  This is repeated 100 times,
generating 100 distances for each of the three methods.  A pilot run
was used to determine MH proposals for $\beta$, which was subsequently
used throughout.  This initial run also indicated that the marginal
posterior distributions for the $\beta$ coefficients corresponding to
$r$, $T_u$ and $T_l$ was tightly straddling zero, prompting us to
discard these predictors from the model.
\begin{figure}[ht!]
\centering
\begin{minipage}{8cm}
\vspace{0.2cm}
\hspace{3cm} (a) \\
\includegraphics[trim=0 20 0 35,scale=0.5]{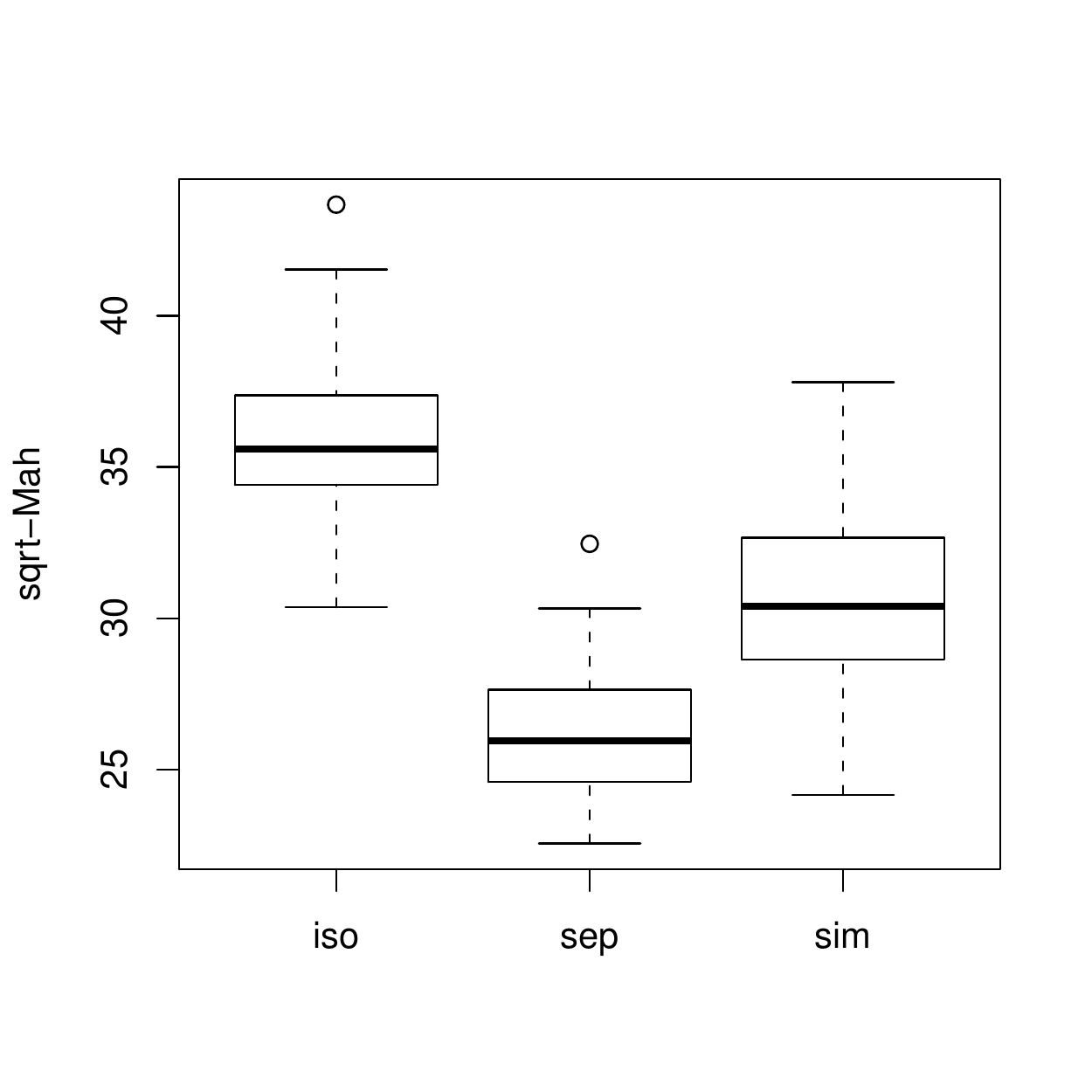}
\end{minipage}
\hspace{0.5cm}
\begin{tabular}{l|rrr}
(b)& \multicolumn{3}{c}{$\sqrt{\mathrm{Mah}}$} \\
&      iso       &       sep     &         sim        \\ 
\hline
 Min.    & 30.37   & 22.55   & 24.17   \\
 1st Qu. & 34.42  & 24.61  & 28.67  \\
 Median  & 35.60  & 25.96 & 30.41  \\
 Mean    & 35.95  & 26.19   & 30.64  \\
 3rd Qu. & 37.36  & 27.64   & 32.65  \\
 Max.    & 43.67   & 32.46   & 37.81 
\end{tabular}
\caption{Square-root Mahalanobis distance results for the borehole
  data in terms of boxplots (a) and a numerical summary (b).}
\label{f:borehole}
\end{figure}

The results are summarized in Figure \ref{f:borehole}.  Briefly, we
see that the GP-SIM model is competitive with the other GPRs on this
data.  On average (and 90\% of the time) it is better than the GPR
with an isotropic covariance function, but worse than the GPR with a
separable one (also 90\% of the time).  So projecting onto the index
is helpful before measuring correlations with a single length-scale
parameter, but having a separate length-scale parameter for each input
direction is more effective.
\begin{figure}[ht!]
\centering
\vspace{0.2cm}
(a) \hspace{4.4cm} (b) \\
\includegraphics[trim=5 0 0 35,scale=0.45]{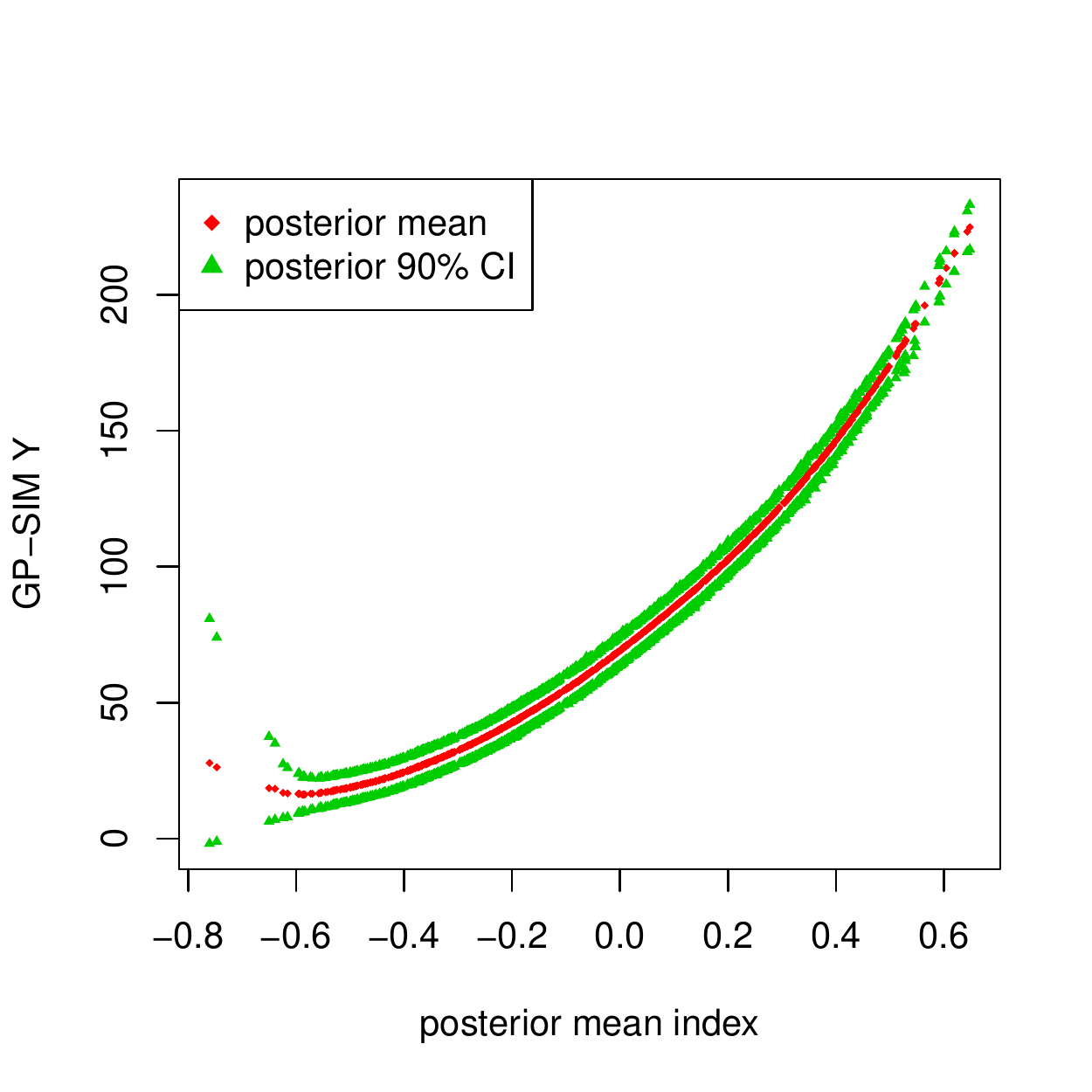}
\includegraphics[trim=20 0 0 35,scale=0.45]{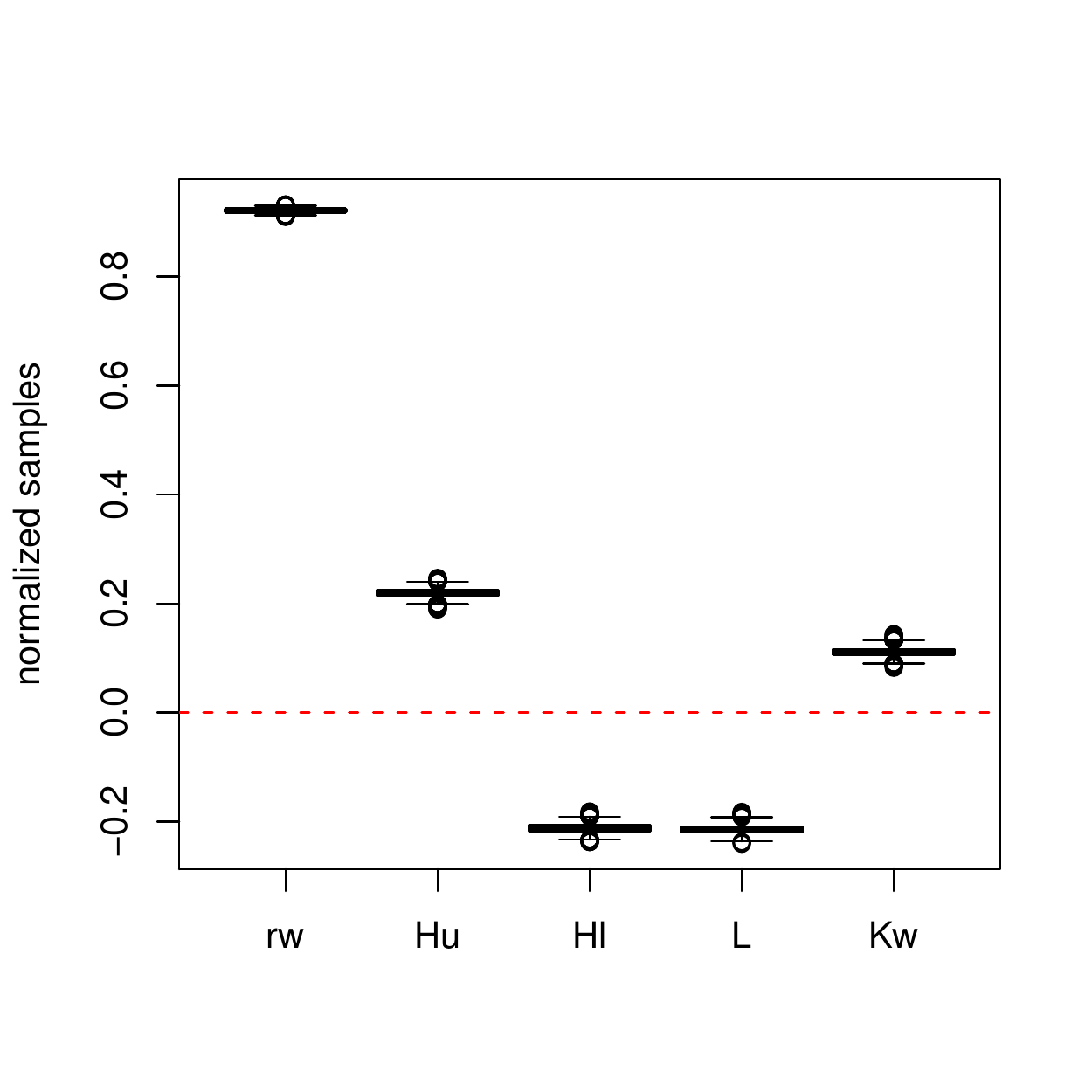}
\caption{Panel (a) shows the posterior mean predicted values, and 90\%
  CI, versus the posterior mean fitted indices; panel (b) shows
  the posterior of the index vector. 
}
\label{f:boreholeindex}
\end{figure}
Figure \ref{f:boreholeindex}(a) shows the posterior distribution of
the responses (mean and 90\% CI) as a function of the posterior mean
indices, describing the contribution of the projection aspect of the
GP-SIM.  Panel (b) shows the normalized estimates of $\beta$. As in
the previous example, the components of $\beta$ from the MCMC had
identical signs throughout, so no post-processing was needed.

In practice, one rarely knows the functional form of the
data-generating mechanism intimately enough to know {\em a priori}
whether an SIM structure (as in Section \ref{sec:synth}) or a
separable structure (as in the current example) is best.  It is
therefore comforting to see that the GP-SIM does not perform
arbitrarily badly with data that are (almost pathologically) outside
of the SIM class. In the next section we show, by example, that
computer experiments can benefit from the estimation of SIM structure
to a surprising degree, especially when one of the inputs plays a
predominant role in predicting the response.

\section{Emulating a real computer experiment}
\label{sec:comp}

To try out the GP-SIM as an emulator for a real computer experiment we
turn to a set of computational fluid dynamics (CFD) codes that
simulate the characteristics of a rocket booster, the Langley
glide-back booster (LGBB), as it is re-entering the atmosphere.  For
previous uses of this data, and further details on the experiment, see
\cite{gra:lee:2009}.  The simulations calculate six aeronautically
relevant responses as a function of three inputs that describe the
state of the booster at re-entry: speed measured in Mach; angle of
attack $\alpha$; and side-slip angle $\beta$, both measured in
degrees.  We shall begin with the {\em roll} response for a detailed
analysis, and revisit the other five responses later.  There are 3014
such quadruplets in the portion of the data set we are concerned with.
Peculiar irregularities (or features) in the relationship between the
inputs and the response, like input-dependent noise and regime
changes, pose challenges for constructing a good emulator and make
this experiment interesting.

First, we wish to see how the GP-SIM measures up against the canonical
GPR models.  Towards this end, we set up an ``inverted'' CV experiment
wherein we partition the data into 10 nearly equal-sized folds.  Then
we iterate over the folds, training the models on the 10\% block of
data inside the fold, and obtaining samples from the posterior
predictive distribution on the remaining 90\% outside the
fold.\footnote{We avoid standard CV since the computational demands
  (required to invert large matrices) would be too large for a Monte
  Carlo experiment.}  Since we do not know the true responses at the
held-out test locations---only the simulated values from the CFD codes
are available---the variance/covariance aspects of the Mahalanobis
distance matrix will play a major role in the comparison.  Rather than
simply penalizing fits which poorly predict a few observations, the
covariance term acknowledges the model's ``explanation'' that they are
noisy.

\begin{figure}[ht!]
\centering
\medskip
\begin{minipage}{9cm}
\vfill
\hspace{4.25cm} (a)\\
\includegraphics[trim=10 50 0 40,scale=0.6]{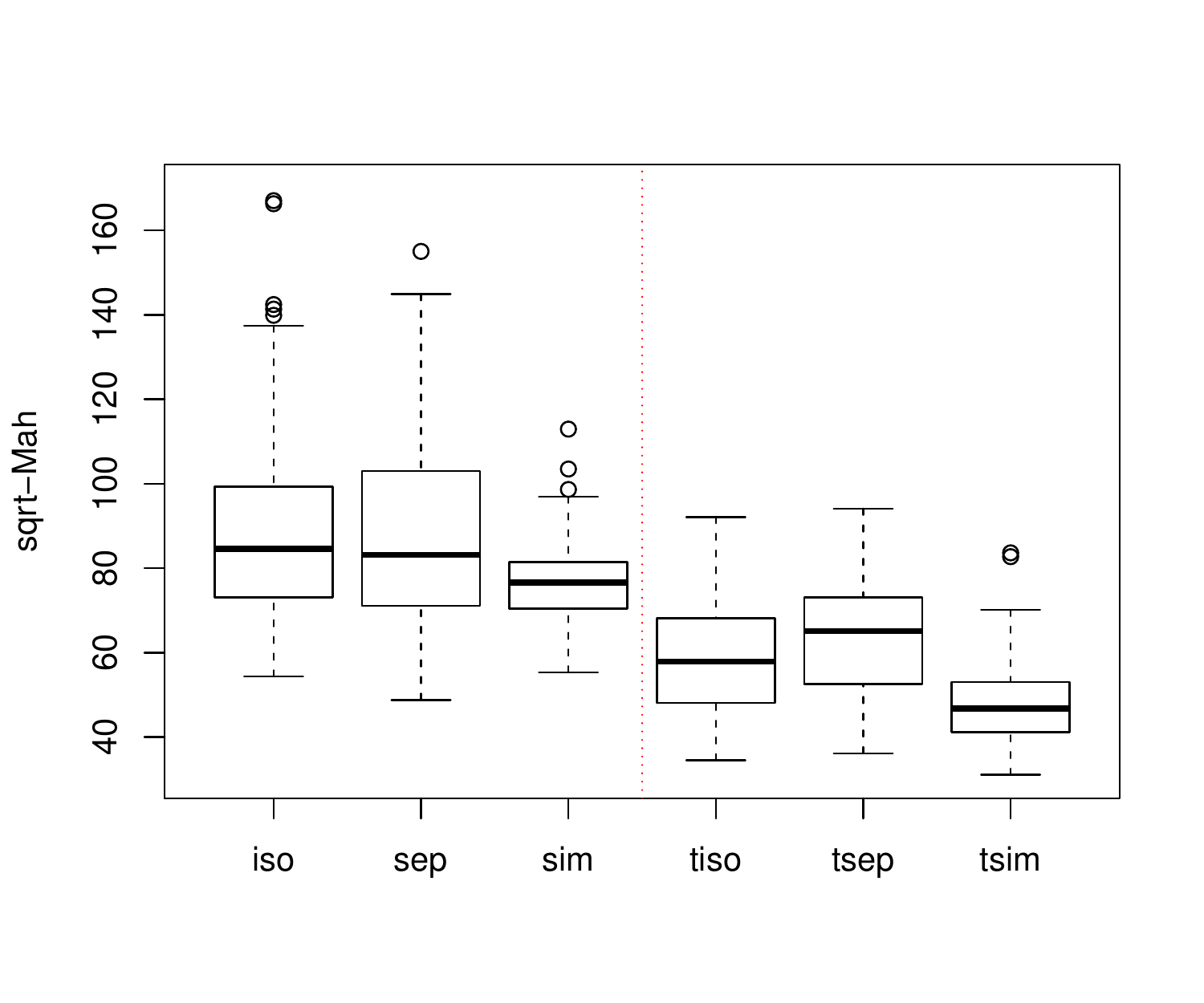}
\vfill
\end{minipage}
\hspace{0.5cm} 
\begin{tabular}{l|rrr}
(b) & \multicolumn{3}{c}{$\sqrt{\mathrm{Mah}}$} \\
&      iso       &        sep         &      sim    \\
\hline
 Min.   & 54.37   & 48.84   & 55.34   \\
 1st Qu.& 73.08   & 71.11  & 70.40  \\
 Median & 84.66   & 83.16 & 76.55   \\
 Mean   & 89.36   & 92.76  & 76.62   \\
 3rd Qu.& 99.27   & 102.98  & 81.40   \\
 Max.   & 166.99   & 256.68 & 112.92 \\
 \hline \\
&    tiso    &       tsep  &         tsim    \\
\hline
 Min.    &34.53   & 36.19  & 31.09  \\
 1st Qu.&48.13   & 52.56  & 41.25  \\
 Median &57.91  & 65.11 & 46.87  \\
 Mean    &59.04  & 64.02 & 47.75  \\
 3rd Qu. &68.21  & 73.05 & 53.08  \\
 Max.     &92.14  & 94.11   & 83.64  
\end{tabular}
\medskip
\caption{Square-root Mahalanobis distances for the LGBB roll response
  in terms of boxplots (a) and a numerical summary (b) for numerical
  specificity. [See Section \ref{sec:tgp} for an explanation of the
  ``t-'' results.]  Observe that $y$-axis of the boxplot clips some of
  the outliers in order to improve visualization. }
\label{f:lgbbmah}
\end{figure}

We applied this inverted-CV procedure ten times, randomly, for 100
total folds generating 100 Mahalanobis distances for GP-SIM and the
two GPR models.  The results are summarized in Figure \ref{f:lgbbmah}.
The figure is actually summarizing the results from two experiments,
the second of which is described later in Section \ref{sec:tgp}.  For
now we focus on the part summarizing fits for ``iso'', ``sep'', and
``sim'' comparators (the {\em left} part of the plot, or the {\em top}
part of the table).  All three versions have similar mean Mahalanobis
distances, although the GP-SIM model was the best on average.  What is
particularly striking from the boxplots is that the variance of the
GP-SIM distances is much smaller than the others, indicating a much
more reliably good fit.  Apparently, projecting the inputs onto a
single index, and measuring spatial correlation on that scale, is
better than estimating axis-aligned spatial covariation (the separable
GPR) or isotropic spatial correlation on the original inputs.

\begin{figure}[ht!]
\centering
\vspace{0.2cm}
(a) \hspace{4.5cm} (b) \\ 
\includegraphics[trim=5 0 0 10,scale=0.45]{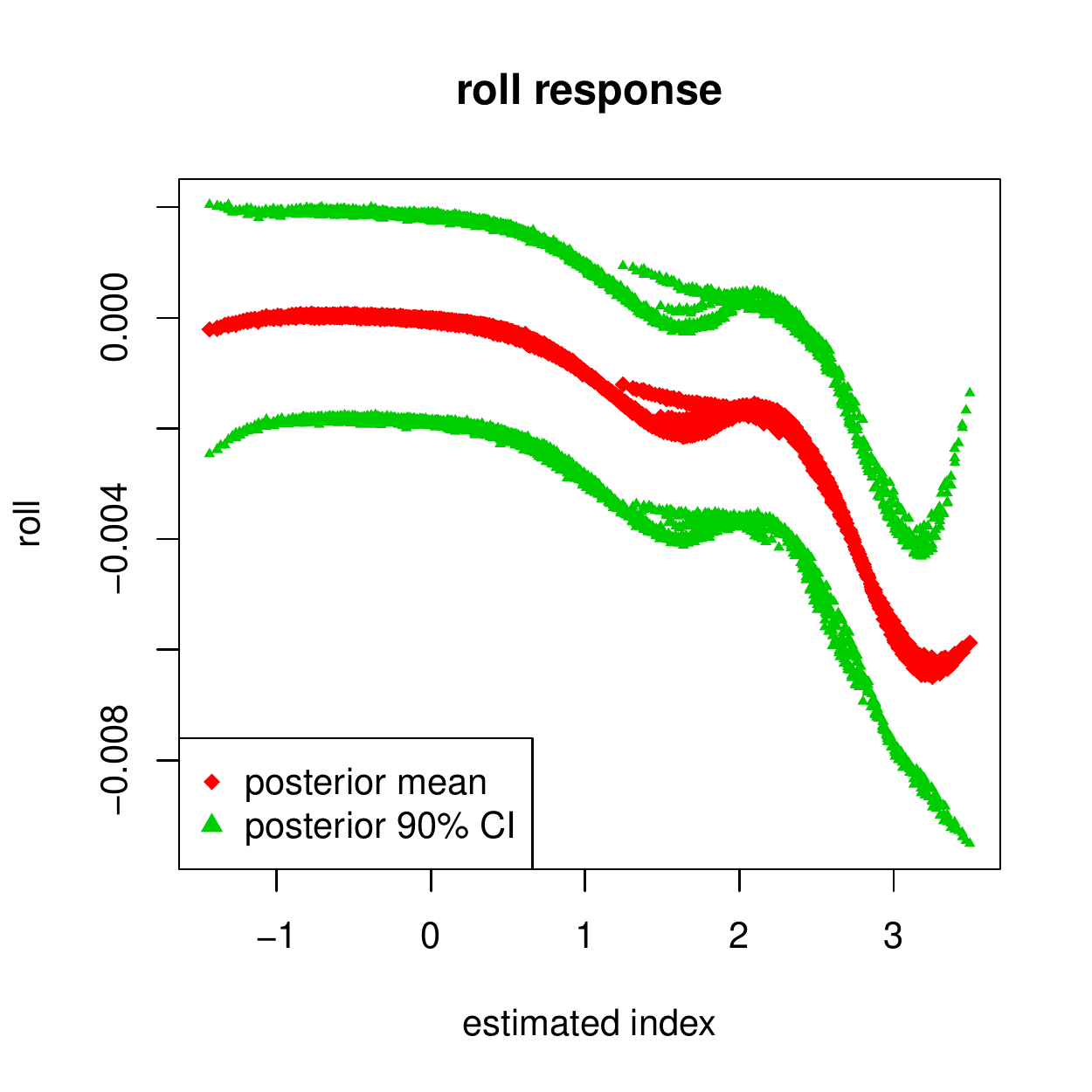}
\includegraphics[trim=20 0 0 10,scale=0.45]{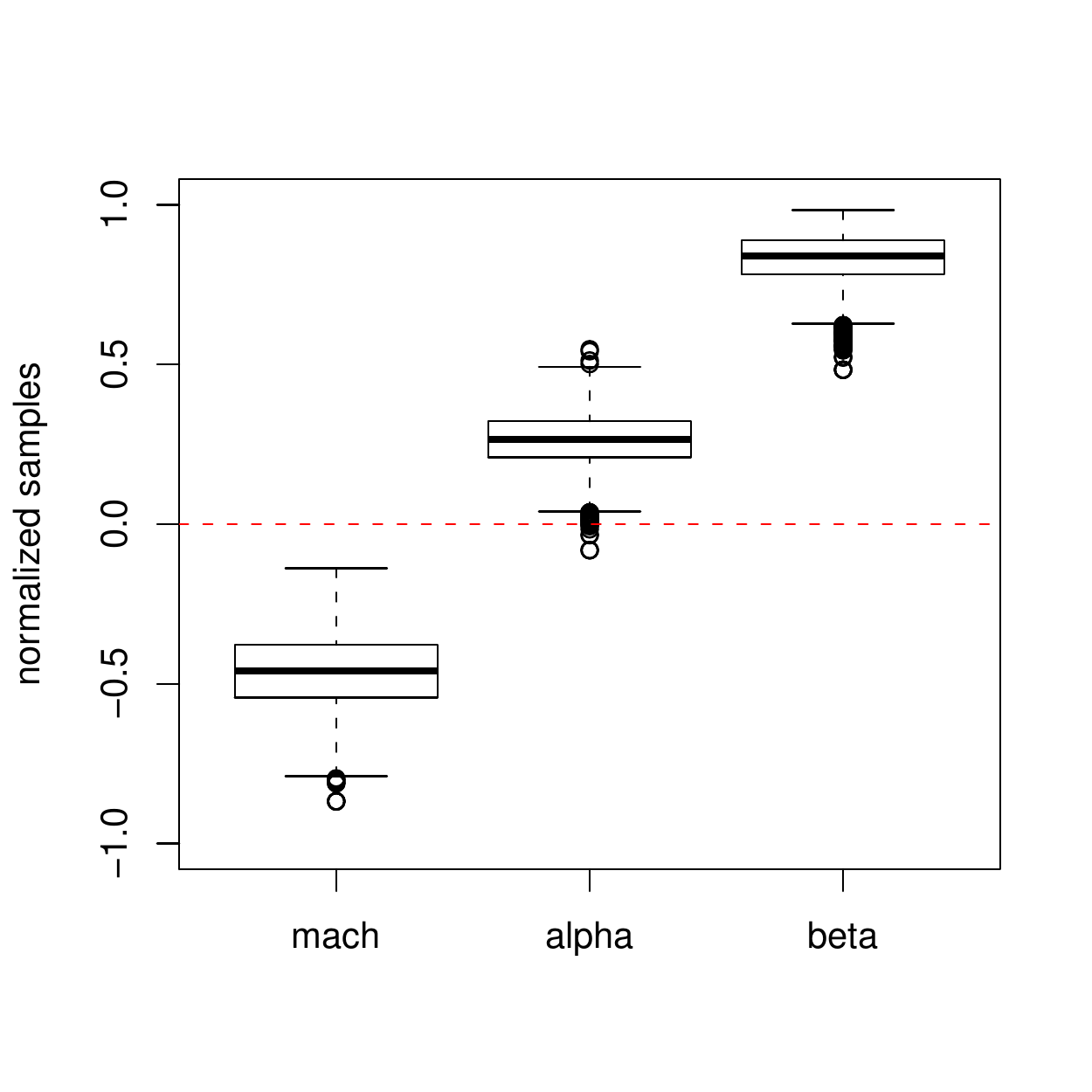}
\caption{Panel (a) shows the posterior mean predicted values versus
  the posterior mean fitted indices; panel (b) shows the posterior of
  the index vector. } 
\label{f:lgbbindex}
\end{figure}
Aspects of the GP-SIM fit are shown in Figure \ref{f:lgbbindex}.  The
curves in panel (a), showing the posterior predictive means and 90\%
CIs versus the posterior mean fitted indices, go some ways towards
explaining why the GP-SIM has lower Mahalanobis distances compared to
the other GPR models.  The non-trivial shape and overall smoothness of
the estimated index--response relationship suggests that the single
index explains a great deal of the variation in the data. An exception
might be for indices in the range $(1.5,2.5)$.  Here disparate inputs,
with similar but distinct input/output relationships, are mapping to
nearby indices and the single index structure is struggling to cope.
A modification to accommodate nonstationarity of the response may help
[see Section \ref{sec:tgp}]. Panel (b) in the same figure shows the
posterior distribution of the components of the index
vector, 
suggesting a reasonable fit with low MC error.  The low variance on
the $\beta$ components with posterior mass far from the origin
suggests that all three inputs are relevant predictors.

\begin{figure}[ht!]
\centering
\includegraphics[trim=5 40 20 45,scale=0.385,clip=TRUE]{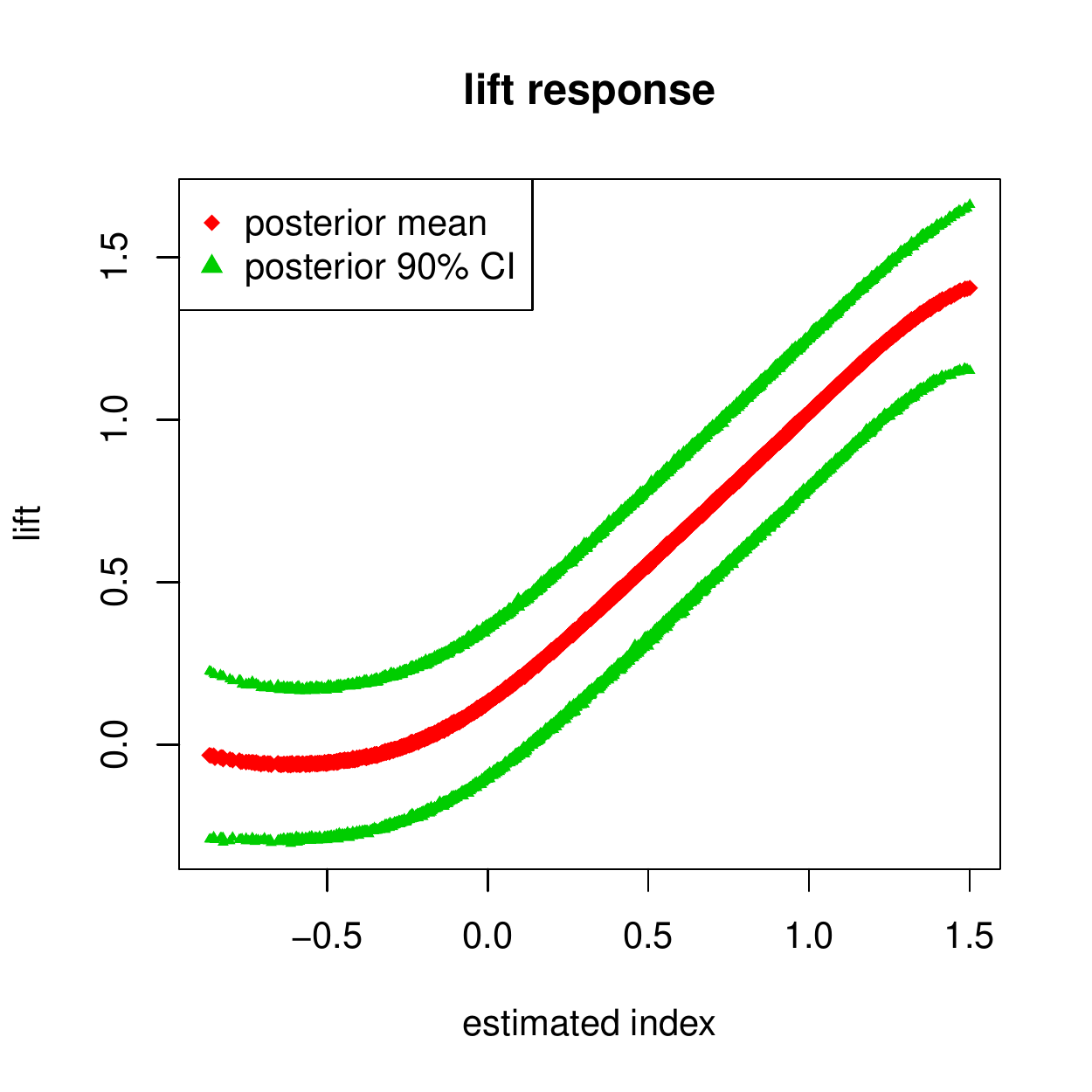}
\includegraphics[trim=0 40 0 45,scale=0.385,clip=TRUE]{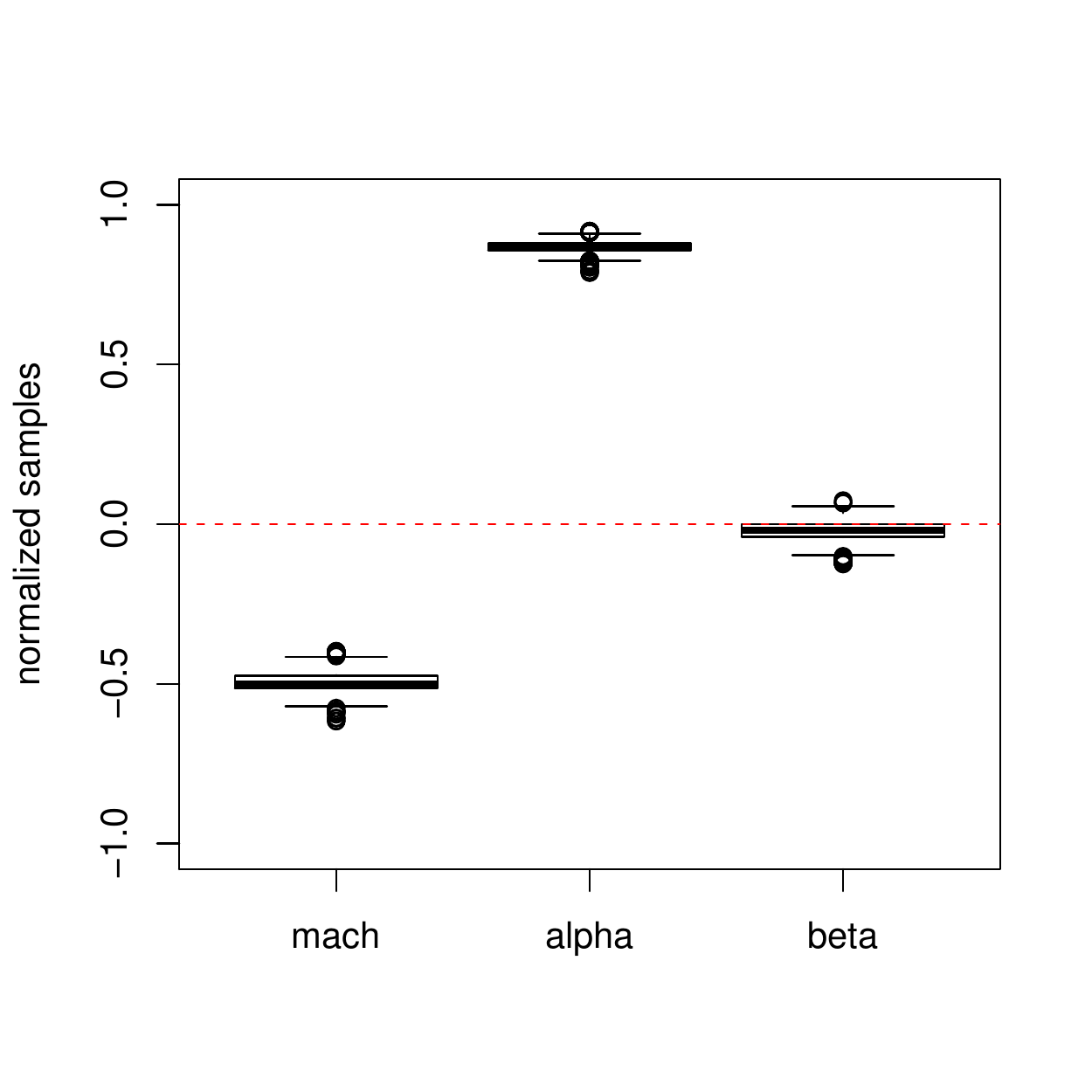}\\
\includegraphics[trim=5 40 20 45,scale=0.385,clip=TRUE]{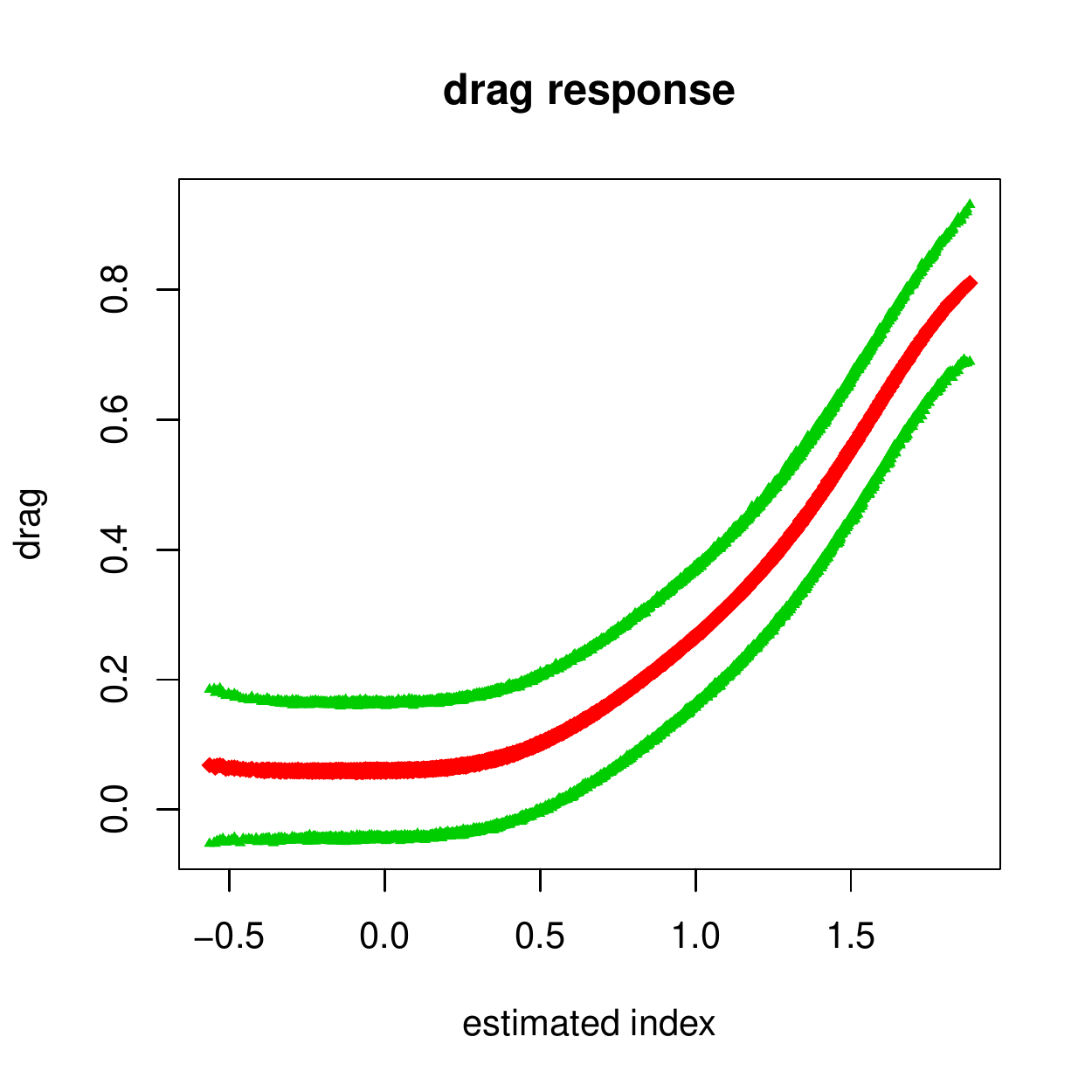}
\includegraphics[trim=0 40 0 45,scale=0.385,clip=TRUE]{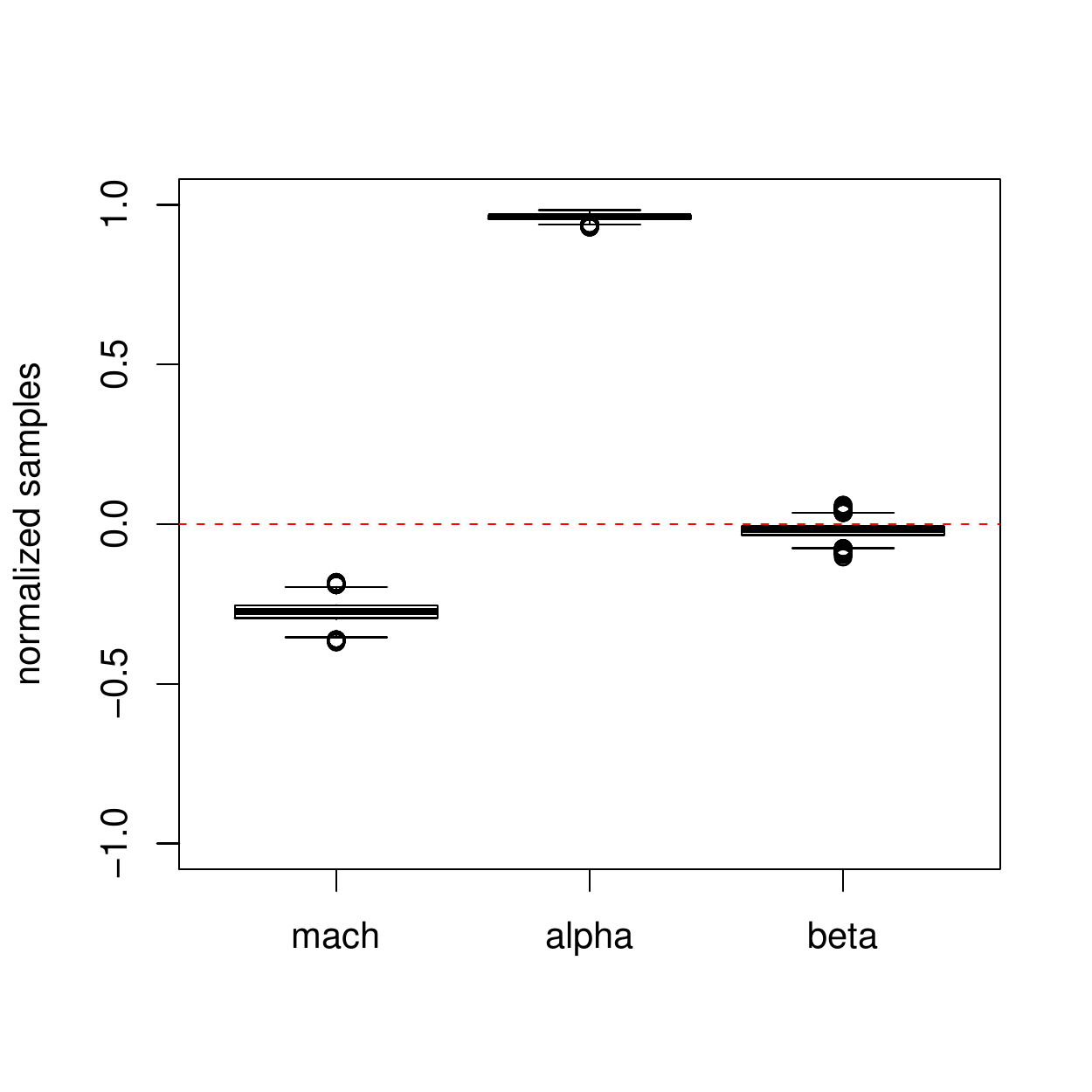}\\
\includegraphics[trim=5 40 20 45,scale=0.385,clip=TRUE]{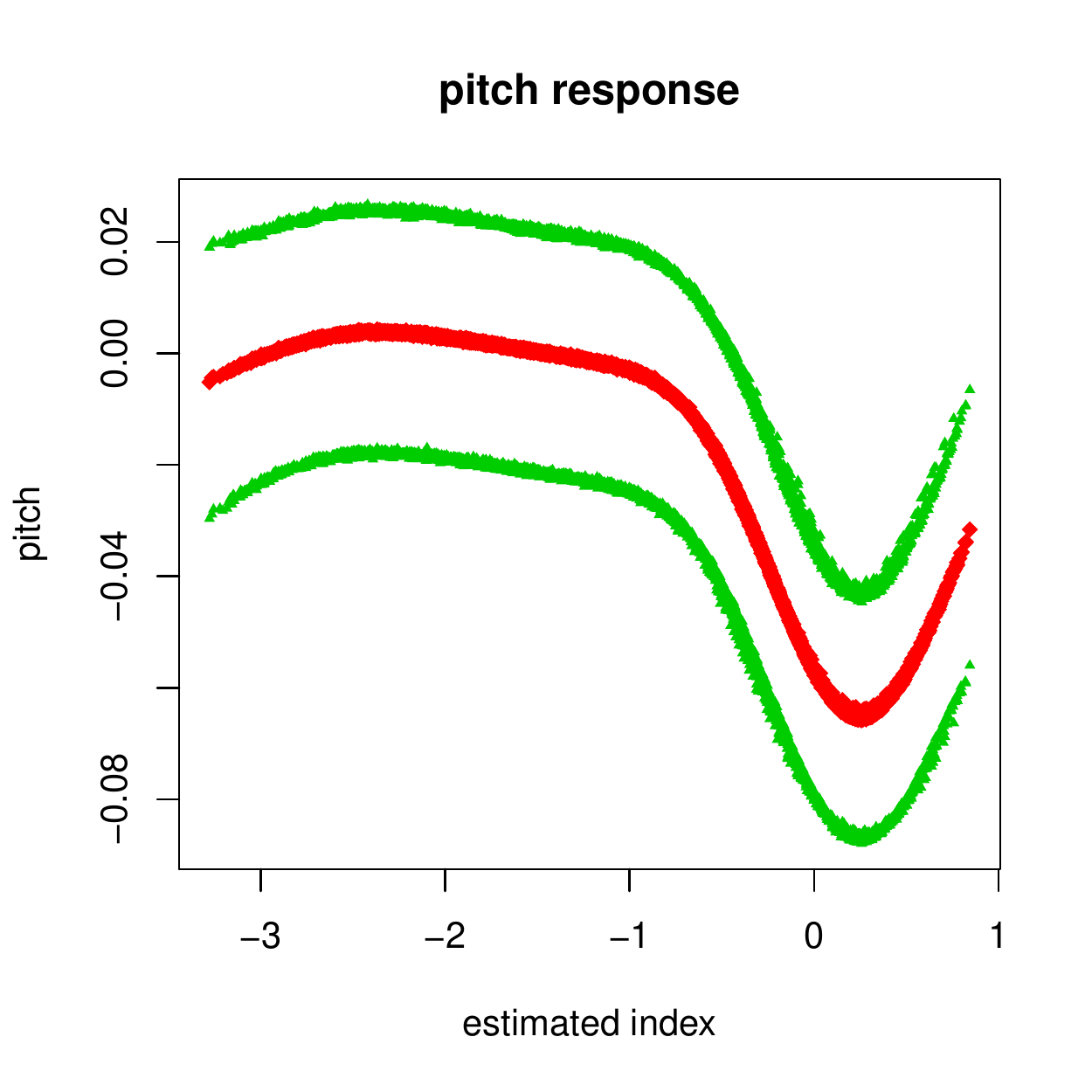}
\includegraphics[trim=0 40 0 45,scale=0.385,clip=TRUE]{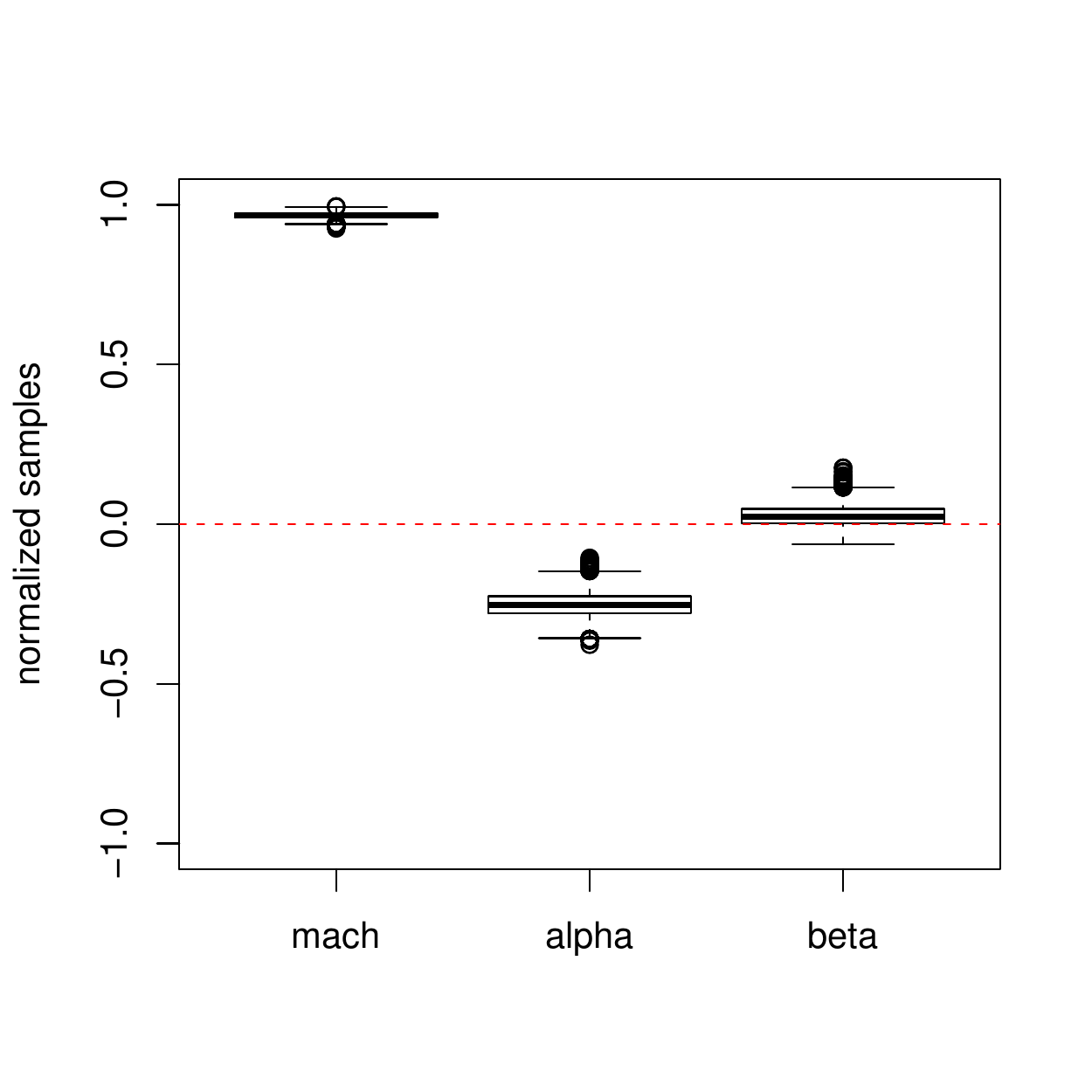}\\
\includegraphics[trim=5 40 20 45,scale=0.385,clip=TRUE]{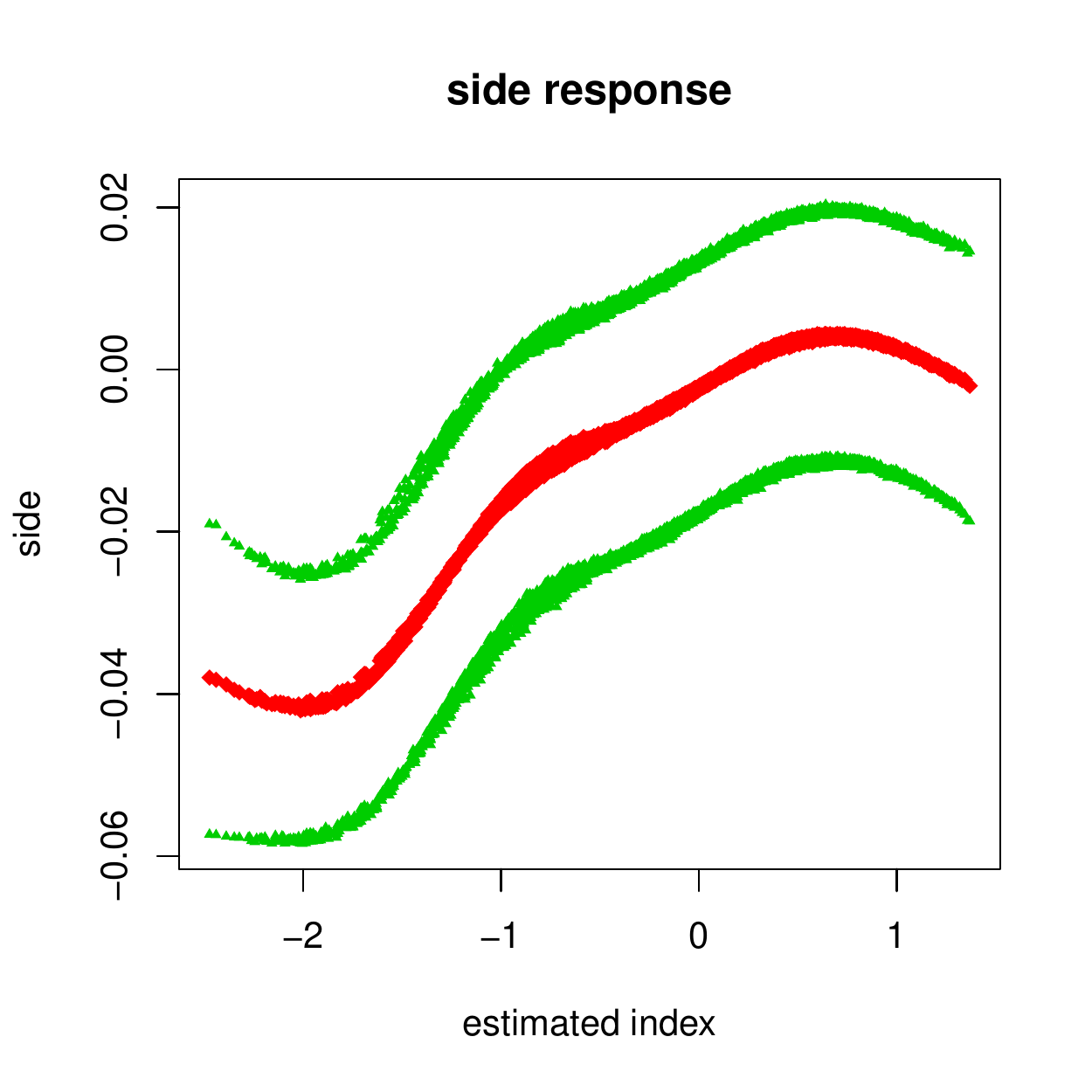}
\includegraphics[trim=0 40 0 45,scale=0.385,clip=TRUE]{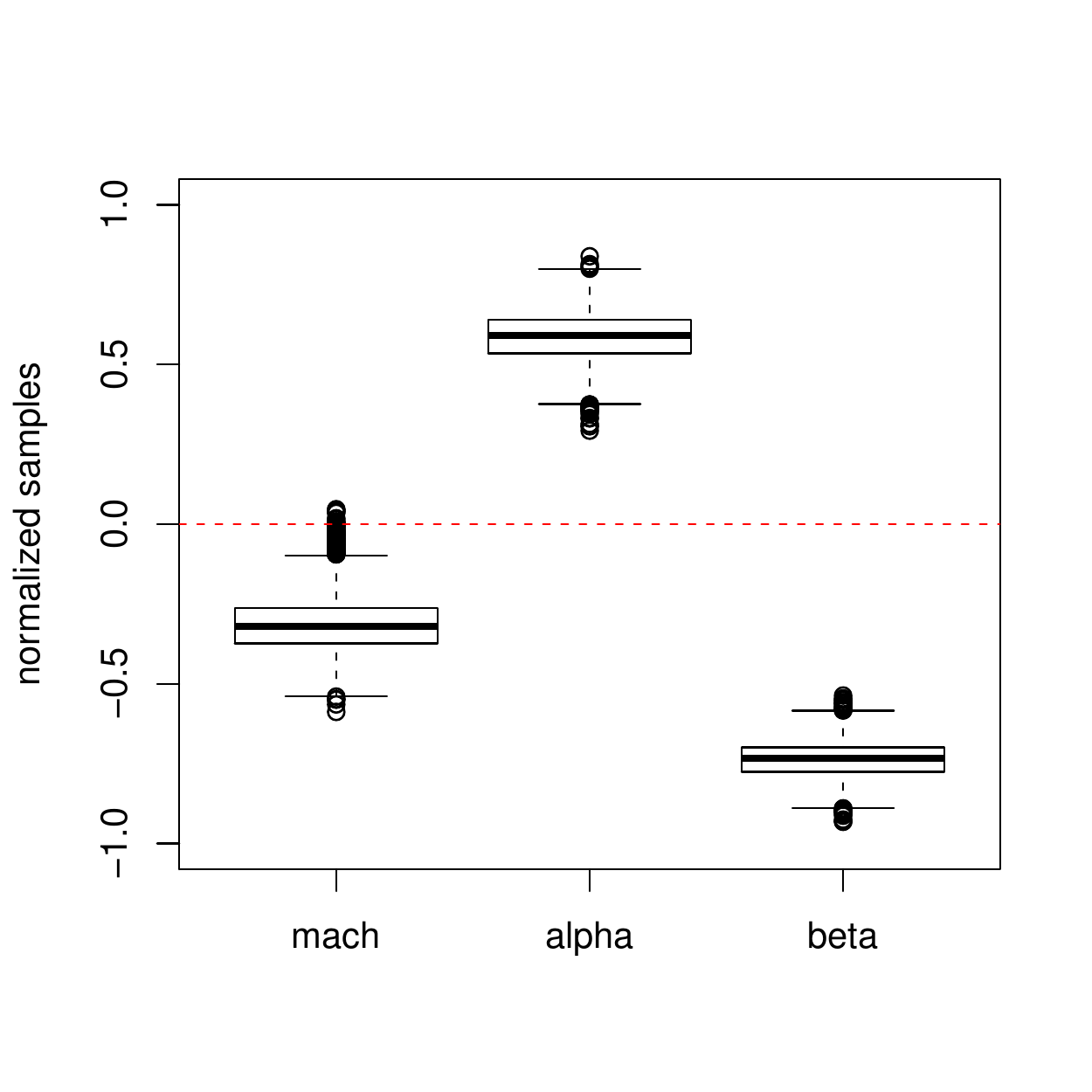}\\
\includegraphics[trim=5 20 20 45,scale=0.385,clip=TRUE]{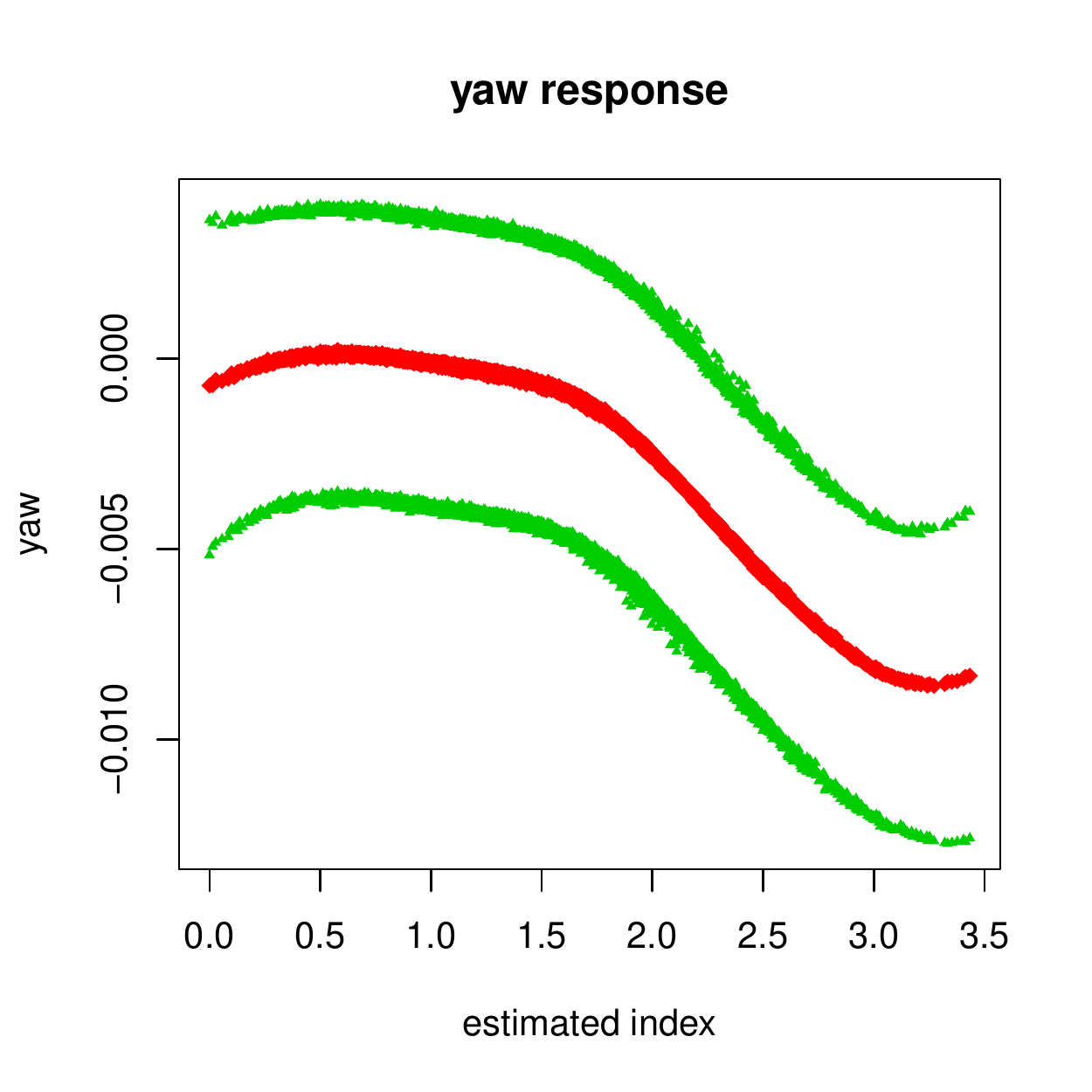}
\includegraphics[trim=0 20 0 45,scale=0.385,clip=TRUE]{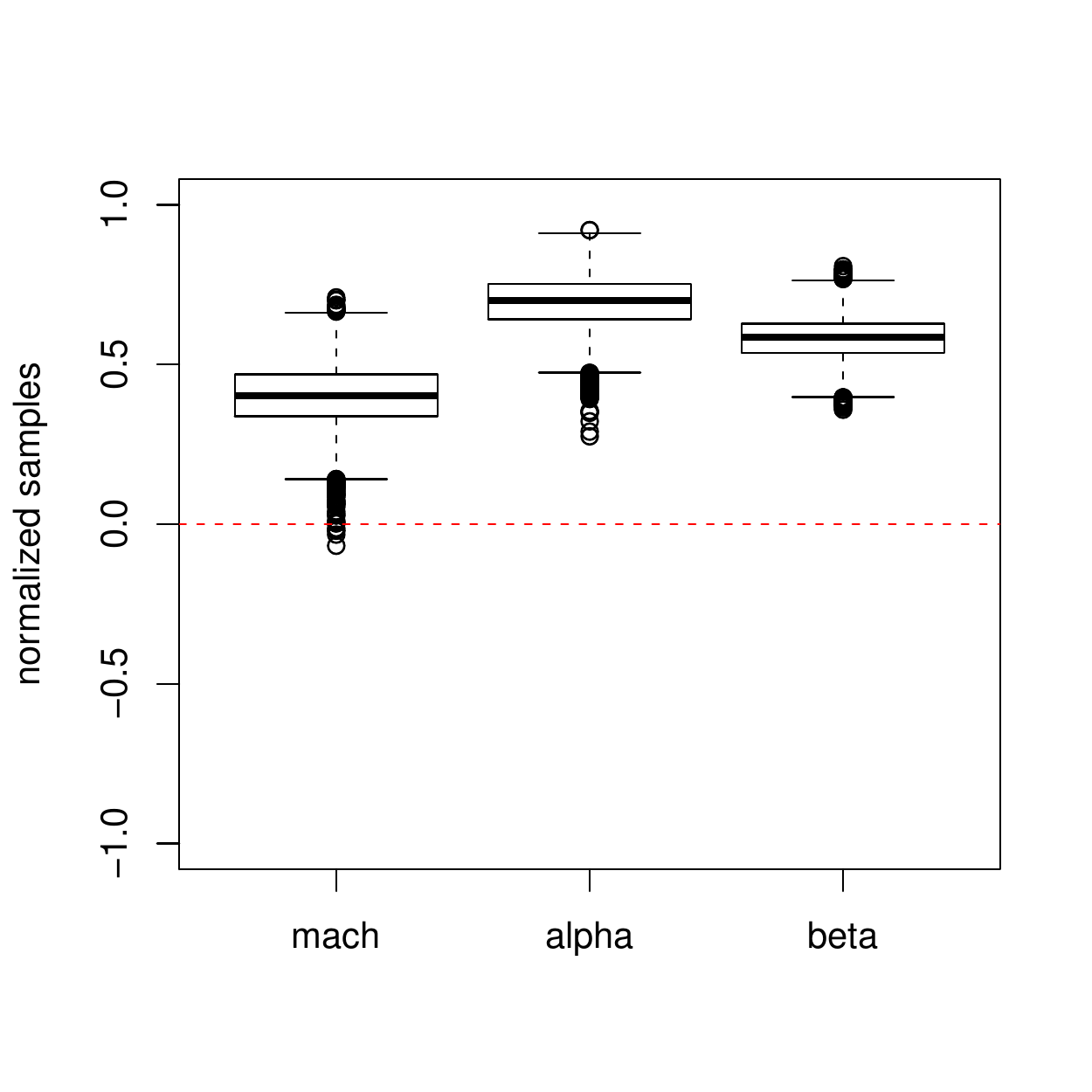}
\caption{{\em Left} column shows the posterior predictive means and
  90\% CI versus the posterior mean indices; {\em right} shows the
  posterior of the index vector.  The rows are {\em lift}, {\em drag},
  {\em pitch}, {\em side}, and {\em yaw}.}
\label{f:lgbbrest}
\end{figure}

The other five responses exhibit broadly similar behavior.  The
summary of the inverted-CV Mahalanobis distances are nearly identical
to those obtained for the {\em roll} response, so they are not
duplicated here.  It is revealing to look at the relationship between
the estimated indices and the response, and the corresponding samples
from the posterior of the index vector, for these other five
responses.  See Figure~\ref{f:lgbbrest}.  The samples of $\theta$ were
similar to the {\em roll} case and so they have been omitted.  Some
brief comments are in order.  The index--response relationship in the
{\em lift} and {\em drag} responses is rather tame, and we can see
that third component of $\beta$, the side-slip angle (also called
$\beta$), is likely not a relevant predictor for these responses---it
tightly straddles zero.  A more formal variable selection analysis may
be warranted \citep{wang:2009}, but is probably overkill in this
particular case of three inputs.  The {\em pitch} index--response
relationship bears some similarity to that for {\em roll}, and like
{\em roll} all of the components of the index vector seem to be
significant.  None of these inputs lead to quite the same
multi-modality of the index--response relationship as in the {\em
  roll} output, suggesting that this case is the most challenging of
the six, and consequently the most likely to benefit from a
nonstationary approach to modeling GP correlations.

To sum up, not only is the GP-SIM a better emulator for this data than
the canonical GPRs, but it offers scope for interpretation and
analysis that is uncanny in the context of computer experiments,
specifically, and GP models generally.

\section{Discussion of extensions}
\label{sec:discuss}

Our reinterpretation of the GP-SIM as a GPR model with a rank-1
covariance function means that the SIM is, now, extremely modular.
The following sub-sections suggest how it can be trivially embedded
into a number of different environments which either extend or
generalize the model, or enable it to be used in a new context.  With
the exception of the last one [Section \ref{sec:mim}], all of these
extensions are {\em already} implemented in one of the {\sf R}
packages, {\tt tgp} \citep{Gramacy:2007,gramacy:taddy:2010} or {\tt
  plgp} \citep{gramacy:polson:2010}, both on CRAN.  In {\tt tgp}
the GP-SIM functionality is invoked by supplying the argument
\verb!cov = "sim"! to the {\tt bgp} fitting routine.  In {\tt plgp}
you specify \verb!cov = "sim"! in the {\tt prior.GP} function.  So
these are more than just suggestions.  The triviality of the changes
required to the packages to add in the GP-SIM functionality,
essentially adding a few extra routines to implement a new covariance
function, is a testament to its newfound modularity.

\subsection{Treed GP-SIM}
\label{sec:tgp}

Some of the challenging aspects of the LGBB experiment, particularly
poor fits from stationary emulators, motivated a new class of models
called the treed Gaussian process \citep[TGP,][]{gra:lee:2008}.  The
idea is to use a Bayesian tree model \citep{chip:geor:mccu:2002} to
infer a partitioning of the space so that independent GPR models could
be fit in different locales or regions of the input space, thus
facilitating nonstationary, input-dependent, modeling.  The result was
a much better emulator for the LGBB simulations and {\em faster} joint
tree--GP MCMC inference compared to GPs alone due to the
divide-and-conquer nature of trees.  The {\tt tgp} package was built
to implement this new methodology.
The implementation of canonical GPR models that we have
been using so far comes as a convenient special case.

Since the GP-SIM is just a special GPR model it enjoys the tree
extensions as well, giving rise to the TGP-SIM.
To entertain the possibility that the TGP-SIM might improve upon the
original TGP model we re-ran the experiment from Section
\ref{sec:comp} using trees.  The results are presented in Figure
\ref{f:lgbbmah} under the headings ``t-iso'', ``t-sep'', and
``t-sim''.  All three methods benefit from the (axis-aligned) treed
partitioning, having lower Mahalanobis distances than their non-treed
counterparts.  This is a testament to the value of the treed
partitioning for this data.  Apparently, projections are the best way
of modeling spatial correlation {\em within} the treed partitions,
leading to a similar ordering as for the non-treed results.
Unfortunately, the interpretation aspects of the GP-SIM model---of
inspecting how the estimated indices relate to the estimated
response---do not so easily translate to the treed version.  So while
the fit is better with trees, the interpretation suffers, as one might
expect from a (much) more nonparametric model.

We also tried the TGP-SIM model on the simple sinusoidal data [Section
\ref{sec:synth}] and the borehole data [Section \ref{sec:borehole}]
but they lead to no improvement.  The tree never partitioned the input
space in either case, so the TGP-SIM model reduced to the GP-SIM
model.  There was also no partitioning for the TGP-GPR models, so they
reduced to canonical GPRs.  Finally, partitioning is a natural way to
handle factor/categorical inputs.  When paired with GPRs the result is
a flexible nonparametric model for regression functions with mixed
categorical and real-valued predictors \citep{brod:gram:2010}.
Therefore the {\tt tgp} implementation represents the first
application of SIM models to mixed inputs [see \citet[][Section
2]{gramacy:taddy:2010} for more details].

\subsection{Sequential design of experiments}
\label{sec:doe}

An important aspect in computer simulation, and consequently an
important application for GPs, is in the designing of the experiment.
Obtaining each response $Y_i$ at input $x_i$ can involve running a
time consuming computer simulation on an expensive machine.  There is
thus interest in minimizing the number of simulations and subsequently
extracting the most information from the experiment.  A common
approach is to design sequentially, using the current model fit to
suggest new $(x,Y)$ pairs by {\em active learning} heuristics.  The
fit is then updated based on the output of the simulation(s) at the
new location(s); and repeat.

Depending on the goal of the experiment some heuristics are more
appropriate than others.  For example, if maximizing understanding
about the $(x,y)$ relationship is the ultimate goal, then statistics
based on the predictive variance or expected reduction in predictive
variance work well \citep{seo00}.  The former is directly available
(\ref{eq:preds2}) and the latter is analytic given the GP
parameterization.  In sequential applications with stationary GPR
models the result is a variation on a space-filling design.  If the
relationship is harder to predict in some parts of the input space
than others, then a model like TGP is more appropriate, and the
resulting sequential designs thus constructed can be far from
space-filling \citep{gra:lee:2009}.  The modularity of the GP-SIM
means that it is easily applied in these contexts.  The {\tt tgp}
package implements both types of active learning heuristic and is
agnostic about the type of correlation (e.g., a rank-1 SIM correlation
can be used).

If optimization---finding $x$ minimizing $Y(x)$---is the goal, then a
statistic called {\em expected improvement}
\citep[EI,][]{jones:schonlau:welch:1998} is appropriate.  This
quantity is also available analytically as a function of the GPR
predictive equations (\ref{eq:predgp}--\ref{eq:preds2}).  Calculations
of EI, and some generalizations, are supported by both {\tt tgp} and
{\tt plgp} thereby further extending the applicability of the GP-SIM
and its treed version.  For more details on EI in the {\tt tgp}
package, see \citet[][Section 4]{gramacy:taddy:2010}.  A particular
generalization of EI for constrained optimization, with the help of
classification models and {\tt plgp}, is discussed below.

\subsection{GP-SIM for classification}

GP models are also popular for classification (GPC).  By using a GPR
model as a prior distribution for latent variables which feed into a
{\em soft-max} function (i.e., an inverse logistic mapping) a
``likelihood'' for categorical responses is implied.  See
\citet{neal:1998}, for details.  Since the GP-SIM is just a special
case of a GPR this means that GP-SIMs can be similarly applied for the
prior distribution on the latent structure, and thus SIMs may be used
for classification too.  GPCs are implemented in the {\tt plgp}
package; 
for details see \citep{gramacy:polson:2010}.  Treed GP classification
\citep{brod:gram:2010} is also possible with the SIM covariance, via
similar extensions.\footnote{The classification extensions to {\tt
    tgp} are not on CRAN, but code may be obtained from the authors.}

A knock-on effect of trivial GP-SIM classification models is that they
can be used in tandem with GP regression models (including SIM) to
solve the sequential design problem of optimization under unknown
constraints.  \cite{gramacy:lee:2010} developed an algorithm that
leverages a statistic called {\em integrated expected conditional
  improvement} by combining a global improvement statistic derived
from a GPR posterior with the probability of satisfying a constraint
from a similar GPC posterior.  This two-pronged approach was
illustrated on a constrained optimization problem arising from computer
simulations in health policy research.  The calculation of these
statistics for all GP combinations, including SIM, is implemented in
the {\tt plgp} package.  As above, invoking this new functionality is
simply a matter of specifying \verb!cov = "sim"! to a function called
\verb!prior.ConstGP!.  In a purely classification context a natural
design heuristic is the predictive entropy, which is also supported
for the CGP-SIM model in the package.

\subsection{Multiple-index models}
\label{sec:mim}

As a natural extension of the SIM, a multiple-index model
\citep[MIM,][]{xia:2008} assumes that all information in the
regression function provided by $X$ is contained in $k$ linear
combinations of the columns of $X$, that is, $\bE\{Y_i|x_i\} =
f(x_i^\top B)$ with $p \times k$ {\em index matrix} $B$. For
identifiability, the constraint $B^\top B=I_k$ is often imposed. Using
our new formulation, it is easy to see that the hierarchical structure
(\ref{eq:hier2}) need not change much to implement a GP-MIM.  The only
substantive difference is that the correlation structure
(\ref{eq:Kstar}) would now have an inverse length-scale parameter of
$\Sigma = BB^\top$, a rank-$k$ matrix.  It is similarly possible to
dispense with the orthogonality constraint although $B$ is then only
identifiable up to an orthogonal $k\times k$ matrix.  Clearly the MCMC
becomes more involved with a higher dimensional parameter like $B$,
necessitating even more care in the design of RW proposal mechanism.
Inference for the rank, $k$, may be facilitated by reversible jump
MCMC \citep{gree:1995} in low-$p$ settings.  Finally, we note that
extending GP-SIM and GP-MIM to employ correlations function other than
the Gaussian case (\ref{eq:K}) is also possible.


\subsection*{Acknowledgments}

RBG is grateful for support from the Kemper Family Foundation.  HL is
supported by Singapore Ministry of Education Tier 1 Grant 36/09.  We
would like to thank to three referees, an associate editor and an
editor whose many constructive comments, collectively, led to a much
improved paper.

\appendix

\section{Heuristics for reconciling the signs}
\label{a:sign}

We see at least three ways of reconciling the signs, or ``labels'' (in
mixture modeling vocabulary), for the index vector, i.e., $\beta$
or $-\beta$.  The first involves looking at the posterior projected
sample indices, whereas the latter two work with the samples of
$\beta$ from the posterior directly.  We note that if the GP-SIM model
is being used solely as a predictive model then there is no need to
identify the labels.  But when aspects of the projection are of direct
interest, heuristics are needed.  This is true under both GP-SIM
formulations discussed in this paper, or indeed any other Bayesian
approach employing a prior distribution for $\beta$ on the (possibly
scaled) unit ball \citep[e.g.,][]{anto:greg:mck:2004}.

Our preferred heuristic involves collecting a set of sample indices
obtained at a reference set of predictive locations $\tilde{X}$
uniformly in $[0,1]^p$, i.e., the collection $\tilde{X} \beta^{(t)}$,
for $t=1,\dots, T$ MCMC samples.  We usually find that the average of
sample indices thus obtained neatly cluster into positive or negative
groups.  The clustering implies a 2-partition of the collection of
$\beta$ samples, one of which has the ``wrong'' sign.  After
``correcting'' the sign of the ``wrong'' group (by negating those
samples) we obtain sample indices which are on one side of zero on
average.  Once so adjusted, plots of average indices and
boxplots/histograms of samples of $\beta$ become are easier to
interpret (and look much like like the ones we provide in this paper).
This technique is certainly not fool-proof.  In particular, it relies
on having a large enough MCMC sample and predictive set $\tilde{X}$ so
that the average indices cluster neatly.

A simpler, but perhaps less reliable or automatic, approach involves
looking at the samples $\beta^{(t)}$ directly.  If some $|\beta_i| \gg
0$ with high posterior probability, say $|\beta_1| \gg 0$, then
reconciling the signs is easy.  Just negate each $\beta^{(t)}$ for
which $\beta_{i1}^{(t)} < 0$, say.  Alternatively, calculate the
covariance matrix of the full sample $\beta^{(1)}, \dots, \beta^{(t)}$
and identify which components of $\beta$ differ in sign via negative
sub-diagonal columns or rows of the matrix.

\begin{figure}[ht!]
\centering
(a) \hspace{4.75cm} (b) \hspace{4.75cm} (c)\\
\includegraphics[scale=0.45,trim=10 0 16 40]{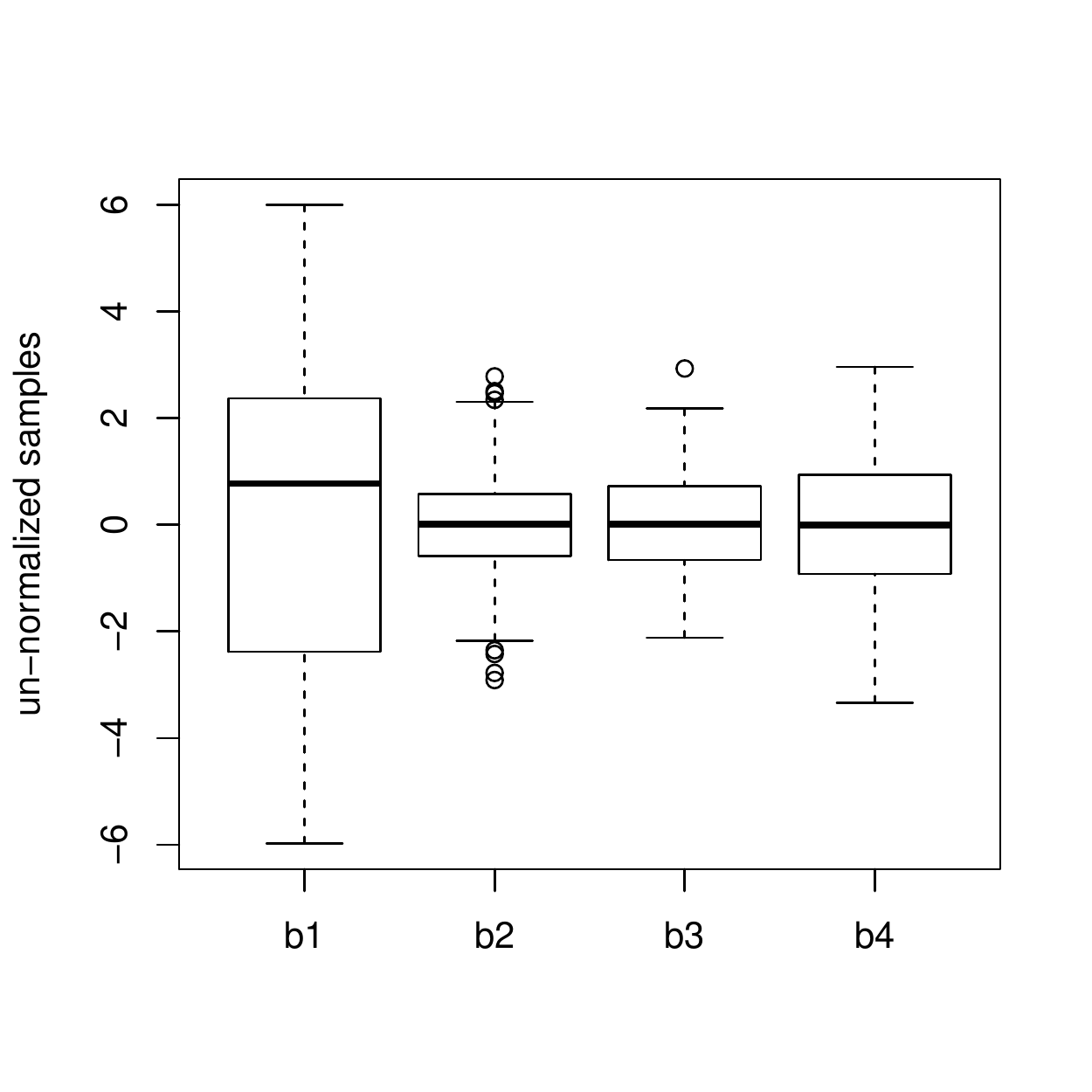}
\includegraphics[scale=0.45,trim=0 0 16 40]{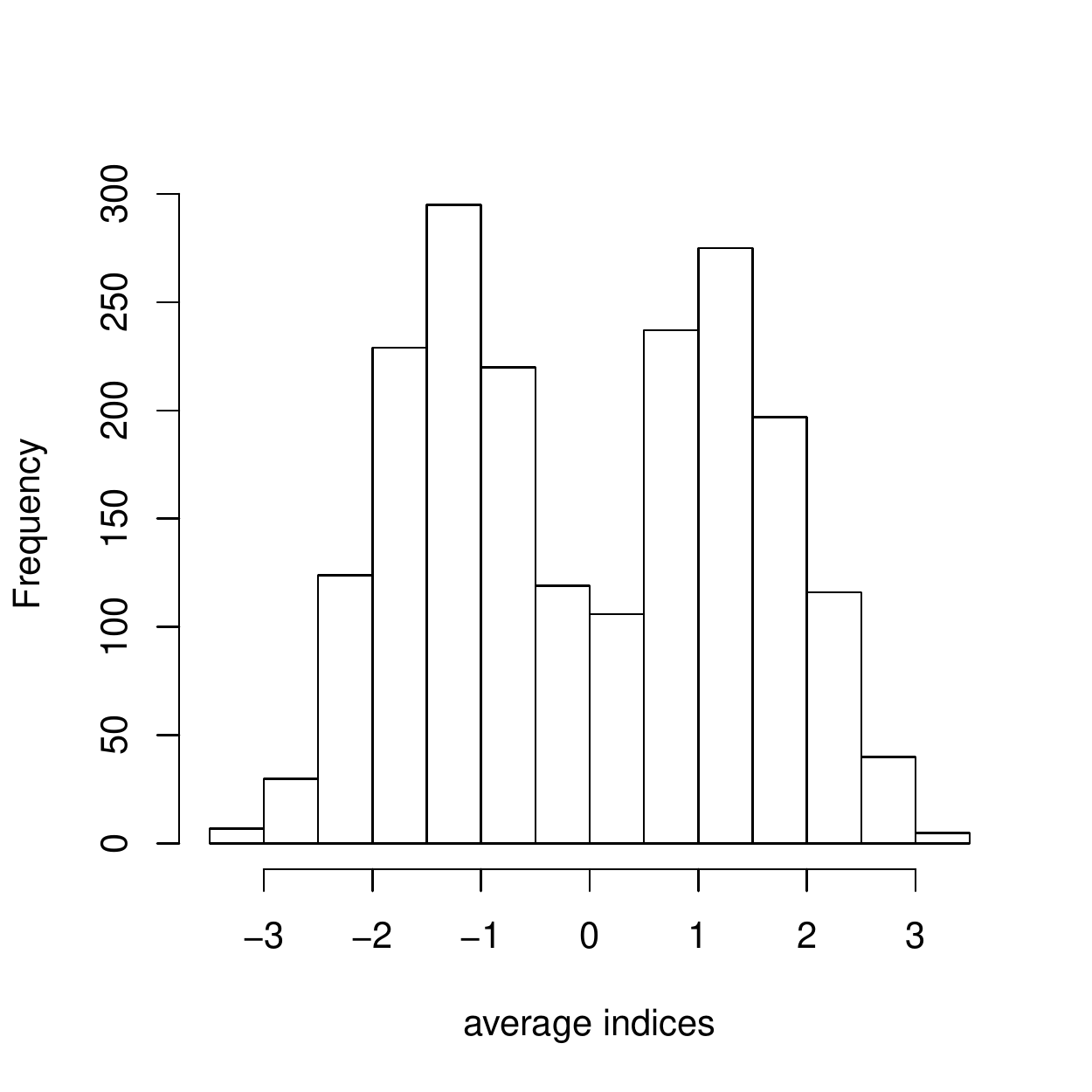}
\includegraphics[scale=0.45,trim=0 0 16 40]{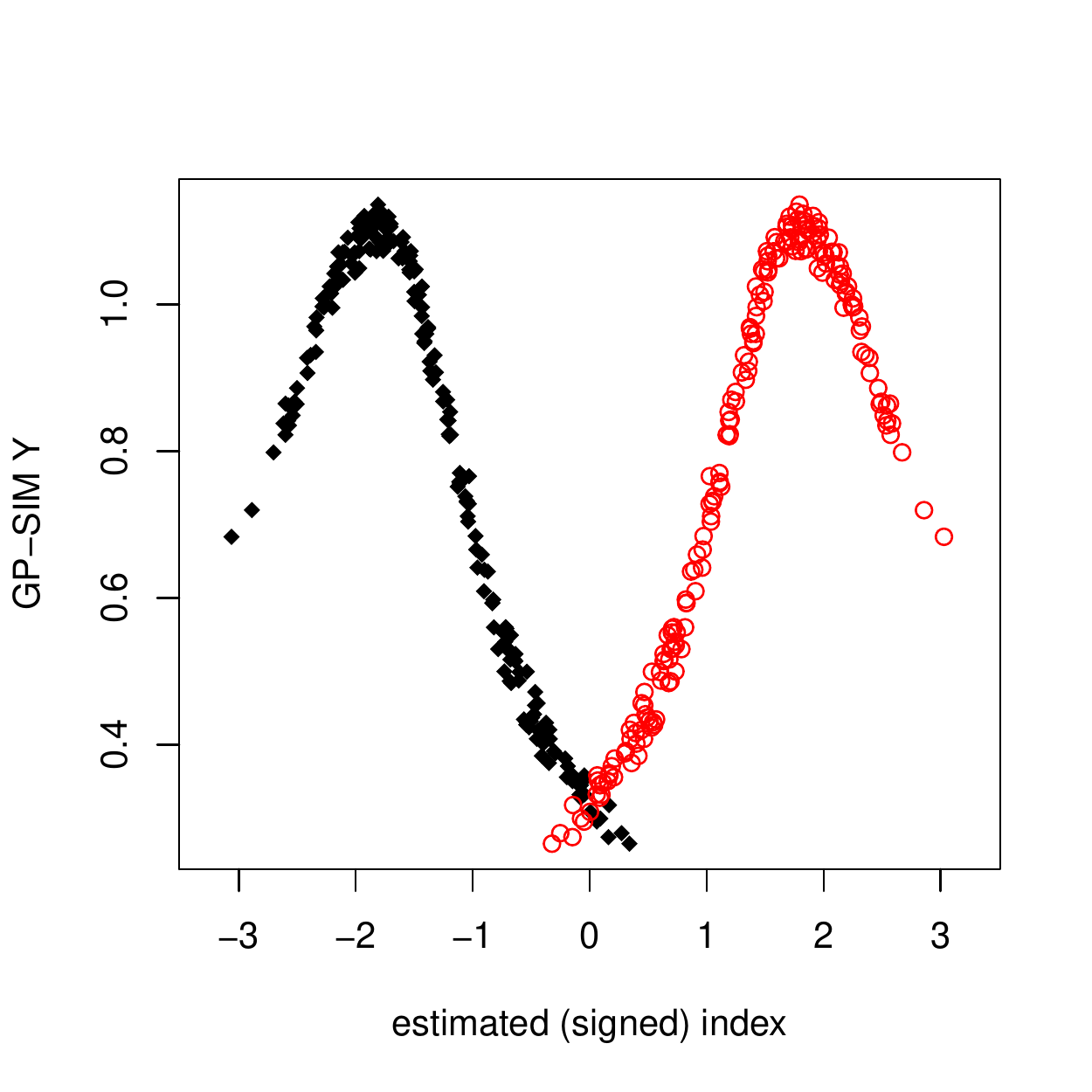}

\vspace{0.25cm}
 (d) ~
\begin{tabular}{r|rrrr}
 &          $\beta_1$     &    $\beta_2$  &  $\beta_3$   & $\beta_4$ \\
\hline
$\beta_1$ & $4.254$ & $1.245$ & $1.997$ & $-1.313$ \\
$\beta_2$ & $1.245$ & $0.374$ & $0.589$ & $-0.384$ \\
$\beta_3$ & $1.998$ & $0.589$ & $0.950$ & $-0.618$ \\
$\beta_4$ &$-1.313$ & $-0.384$ & $-0.618$ & $0.414$
\end{tabular}
\vspace{0.25cm}
\caption{Illustrating the index--based heuristic [{\em top} row:
  panels (a--c)], and the covariance heuristic [{\em bottom} row: panel
  (d)].  Panel (a) shows the sampled (signed) $\beta$s from the
  posterior via boxplots; panel (b) shows the average indices obtained
  from that sample; panel (c) shows the clustered posterior mean
  index--response relationship it suggests; and panel (d) shows the
  posterior covariance matrix of the $\beta$s.}
\label{f:signs}
\end{figure}

Figure \ref{f:signs} illustrates these two methods on the sinusoidal
synthetic data from Section \ref{sec:synth}.  The {\em top} row,
panels (a--c), show how the method based on sample indices would play
out.  Although the sample $\beta$s are both positive and negative
[panel (a)], panel (b) shows that the indices cluster nicely.  Panel
(c) shows the implied index--response relationship, where
colors/points indicate which points are in which cluster according the
parity of the average indices.  The {\em bottom} row, panel (d), shows
the posterior covariance matrix of the samples of $\beta$ [shown in
panel (a)], indicating that $\beta_4$ has an opposite sign from the rest.
Using this heuristic leads to an identical clustering to the one shown
in panels (b--c).

The last heuristic involves a different, perhaps more reliable,
approach to finding a point estimator for $\beta$ given samples from
the posterior.  We first normalize these vectors so that
$\|\beta^{(t)}\|=1$.  Because of the sign indeterminacy, we cannot
use the sample average as an estimator.  Instead, we define
$\hat{\beta}=\min_{\beta:\|\beta\|=1}\sum_{t=1}^T
(1-(\beta^\top\beta^{(t)})^2)$. For interpretation of this measure,
note that $1-(\beta^\top\beta^{(t)})^2=\sin^2\theta^{(t)}$ where
$\theta^{(t)}$ is the angle between $\beta$ and $\beta^{(t)}$.  It is
equal to zero for both $\theta^{(t)}=0$ and
$\theta^{(t)}=\pi$. Minimizing $\sum_{t=1}^T
(1-(\beta^\top\beta^{(t)})^2)$ is the same as maximizing $\sum_{t=1}^T
(\beta^\top\beta^{(t)})^2=\beta^\top(\sum_t\beta^{(t)}\beta^{(t)\top})\beta$
and we easily see that $\hat{\beta}$ is just the eigenvector
corresponding to the largest eigenvalue of the matrix
$\sum_t\beta^{(t)}\beta^{(t)\top}$.  This estimator can be used on its
own, or to help ``choose'' a set of signs for the full sample.

\bibliography{gpsim}
\bibliographystyle{jasa}

\end{document}